\documentclass[%
 aip,
 amsmath,amssymb,
reprint,
 jcp,
]{revtex4-1}

\usepackage{graphicx}
\usepackage{dcolumn}
\usepackage{bm}

\usepackage[utf8]{inputenc}
\usepackage[T1]{fontenc}
\usepackage{mathptmx}
\usepackage{etoolbox}

\usepackage{amsmath,amssymb}
\usepackage{float}
\usepackage{subcaption}
\graphicspath{ {images/} }
\usepackage{multirow}
\usepackage{siunitx}

\makeatletter
\def\@email#1#2{%
 \endgroup
 \patchcmd{\titleblock@produce}
  {\frontmatter@RRAPformat}
  {\frontmatter@RRAPformat{\produce@RRAP{*#1\href{mailto:#2}{#2}}}\frontmatter@RRAPformat}
  {}{}
}%
\makeatother
\begin{document}

\preprint{AIP/123-QED}


\title[ML potentials with explicit electrostatic interactions]{Incorporating Coulomb interactions with fixed charges in Moment Tensor Potentials and Equivariant Tensor Network Potentials}
\author{Dmitry Korogod}
\email{korogod.dv@phystech.su}
\affiliation{Skolkovo Institute of Science and Technology, Skolkovo Innovation Center, Bolshoy boulevard 30, Moscow, 143026, Russian Federation}
\affiliation{Moscow Institute of Physics and Technology, Institutsky lane 9, Dolgoprudny, Moscow region, 141700, Russian Federation}
\affiliation{Digital Materials LLC, Odintsovo, Kutuzovskaya str. 4A Moscow region, 143001, Russian Federation}

\author{Olga Chalykh}
\affiliation{Skolkovo Institute of Science and Technology, Skolkovo Innovation Center, Bolshoy boulevard 30, Moscow, 143026, Russian Federation}

\author{Max Hodapp}
\affiliation{CD Laboratory for Digital material design guidelines for mitigation of alloy embrittlement, Materials Center Leoben Forschung GmbH (MCL), Leoben (AT)}

\author{Nikita Rybin}
\affiliation{Skolkovo Institute of Science and Technology, Skolkovo Innovation Center, Bolshoy boulevard 30, Moscow, 143026, Russian Federation}
\affiliation{Digital Materials LLC, Odintsovo, Kutuzovskaya str. 4A Moscow region, 143001, Russian Federation}

\author{Ivan S. Novikov}
\affiliation{HSE University,         Faculty of Computer Science, Pokrovsky boulevard 11, Moscow, 109028, Russian Federation}
\affiliation{Skolkovo Institute of Science and Technology, Skolkovo Innovation Center, Bolshoy boulevard 30, Moscow, 143026, Russian Federation}
\affiliation{Moscow Institute of Physics and Technology, Institutsky lane 9, Dolgoprudny, Moscow region, 141700, Russian Federation}
\affiliation{Emanuel Institute of Biochemical Physics of the Russian Academy of Sciences, 4 Kosygin Street, Moscow, 119334, Russian Federation}

\author{Alexander V. Shapeev}
\affiliation{Skolkovo Institute of Science and Technology, Skolkovo Innovation Center, Bolshoy boulevard 30, Moscow, 143026, Russian Federation}
\affiliation{Digital Materials LLC, Odintsovo, Kutuzovskaya str. 4A Moscow region, 143001, Russian Federation}

\date{\today}

\begin{abstract}
In this work, we incorporate long-range electrostatic interactions in the form of the Coulomb model with fixed charges into the functional form of short-range machine-learning interatomic potentials (MLIPs), particularly in the Moment Tensor Potential and Equivariant Tensor Network potential. We show that explicit incorporation of the Coulomb interactions with fixed charges leads to a significant reduction of energy fitting errors, namely, more than four times, of short-range MLIPs trained on organic dimers of charged molecules. Furthermore, with our long-range models we demonstrate a significant improvement in the prediction of the binding curves of the organic dimers of charged molecules. Finally, we show that the results calculated with MLIPs are in good correspondence with those obtained with density functional theory for organic dimers of charged molecules.
\end{abstract}

\maketitle

\section{Introduction}

Atomistic simulation is a widely used methodology for the theoretical investigation of chemical systems (see, e.g., Ref.~\onlinecite{MD_MLIPs_REVIEW}). An interatomic potential is the core of atomistic simulation. Such a potential enables predicting energies and forces acting on atoms. Machine-learning interatomic potentials (MLIPs) have become one of the most popular classes of interatomic potentials due to their ability to accurately approximate results of calculations performed using first principles methods like density functional theory (DFT) at a fraction of the computational cost.

Most of the developed MLIPs \cite{NNP,GAP,SOAP,SNAP,MTPmulti,DeePMDsr,ACE,PINN,NeQUIP,ETN} are local in the sense that energies predicted by them are sums of contributions of individual local atomic environments of a given cutoff radius around the central atom. Because of locality, these MLIPs do not explicitly capture long-range interactions, and, therefore, cannot accurately predict properties of many systems, including those with dominant electrostatic interactions. Furthermore, local MLIPs are unable to predict partial charges of atoms, which can be used to calculate properties, such as dielectric constants, ionic conductivity, and capacitance. Finally, short-range models can not be used for investigating charged molecules or even their interaction with neutral molecules, as they are not able to correctly predict the potential energy surface which dramatically affects a wide range of physical and chemical properties.

One of the first attempts to include electrostatic interactions in MLIPs involved adding environment-dependent charges to Neural Network Potentials (NNPs)~\cite{NNP}. This approach, termed 3G-HDNNP (Third-Generation High-Dimensional NNP)~\cite{3GHDNNP}, was tested on zinc oxide clusters. However, this model is not capable of capturing global charge redistribution and correctly predicting the total charge of the system. To address these shortcomings of 3G-HDNNP, it was proposed to use the charge equilibration (QEq) scheme~\cite{QEq} in which partial charges are determined during the minimization of the global energy of electrostatic interaction in a system, depending on the electronegativity and hardness of atoms. These parameters were chosen to be tunnable, the hardness based only on the types of atoms and the electronegativity depending on the atomic environments~\cite{CENT}. This scheme is the basis of 4G-HDNNP~\cite{4G-HDNNP}, which demonstrated the ability to capture global charge redistribution. The 4G-HDNNP model correctly predicted the position and adsorption energy of a gold atom on an undoped and fluorine-doped magnesium oxide substrate, whereas 3G-HDNNP failed to do it. The latest stage of NNP developments is the so-called ee4G-HDNNP (Electrostatically Embedded 4G-HDNNP)~\cite{ee4GHDNNP}, in which the short-range part of the energy depends not on the charges, but on the electrostatic potential induced by these charges, and there is an additional term describing dispersion interactions and short-range Pauli repulsion. These changes enabled for more accurate predictions of the relative energies of neutral and charged sodium chloride clusters and significantly improved the transferability of the potential: the model trained on sodium chloride clusters accurately predicted the equation of state of a sodium chloride crystal, unlike 4G-HDNNP.

Besides NNPs, there are other potentials that explicitly include electrostatic interactions. The developers of DeePMD \cite{DeePMDlr} subtracted the electrostatic (Coulomb) energy from the total DFT energy, fitted the short-range DeePMD model to the non-electrostatic energy, and then conducted simulations with the combined DeePMD and Coulomb models. It was demonstrated that DeePMD+Coulomb gives an order of magnitude lower fitting errors than pure DeePMD on the example of water. The DeePMD+Coulomb also outperformed pure DeePMD when computing vapor density and intermolecular potential energy surface of two water molecules. In the DPLR model~\cite{DPLR}, which is based on the DeePMD model, two types of charged sites were introduced instead of partial atomic charges: ionic sites, representing nuclei and core electrons, and electronic sites, representing valence electrons. While the former are simply located on atoms, position of the latter is defined through the averaging of the maximally localized Wannier centers~\cite{WannierCenter} of atoms and predicted using local machine-learning models. This addition improved the accuracy of the binding curve prediction for water molecules of a standard DeePMD model. In Ref.~\onlinecite{LODE} the authors combined the short-range Smooth Overlap of Atomic Positions (SOAP) model with the long-distance equivariant (LODE) framework which uses local descriptors to encode the Coulomb and other asymptotic decaying potentials (1/$r^p$) around the atoms, and a related, density-based long-range descriptor. The SOAP+LODE model correctly reproduced the binding curves of six charged dimers. Recently, Latent Ewald Summation (LES) was combined with the Cartesian Atomic Cluster Expansion \cite{CACE}. In the CACE+LES model \cite{LatentEwald}, partial charges of atoms are hidden multidimensional variables, the interaction energies of which are calculated independently and then summed up. This model improved the accuracy of the short-range CACE model for predicting the distribution of water molecule orientations at the liquid-gas interface. Also, CACE+LES correctly predicted the asymptotics of the interaction of neutral and charged molecules. Finally, in Ref.~\onlinecite{MTP+QEq} explicit electrostatic interactions in the form of the QEq model \cite{QEq} were added to the Moment Tensor Potential (MTP) \cite{MTPmulti}. It was shown that the lack of fitting of the MTP+QEq model to the DFT partial charges and including the electronegativity and hardness depending only on atomic types did not outperform pure MTP on the example of the silica system. 

In this work, we develop a combination of a local MLIP and non-local Coulomb model with fixed partial charges. We take MTPs and Equivariant Tensor Network (ETN) potentials \cite{ETN} as local MLIPs. To parameterize MTPs and ETNs, we train them on energies of structures and forces acting on atoms obtained in the scope of DFT, and we additionally use partial charges from DFT to fit the combined models. We note that the developed models incorporate electrostatic (Coulomb) interactions with charge redistribution (labeled as QRd from now on), which enables preserving the total charge of a system. We demonstrate that explicitly including Coulomb interactions in the functional form of potentials is critical for an accurate prediction of the binding curves of different dimers containing only charged molecules. As opposed to the results in Ref.~\onlinecite{MTP+QEq}, even simple fixed partial charges in the Coulomb model substantially improve local potentials. Further, we show that such a simple charge prediction technique is not expressive enough to significantly improve the short-range model for the dimers containing one charged and one neutral molecule. We show that increasing the cutoff radius is not a solution to this issue for some of the considered dimers.


This work is organized as follows. In the Methodology section, we introduce the two short-range models (MTP and ETN) and the combined models incorporating long-range electrostatic interactions: MTP+QRd and ETN+QRd. We also describe the procedures for fitting these models and for optimizing the ETN hyperparameters. In the Results section, we give the computational details and describe the training sets. To test our methodology, we consider six dimers in vacuum: CH$_3$COO$^{-}$+CH$_3$COO$^{-}$, CH$_3$COO$^{-}$+CH$_3$NH$_3^+$, $\rm C_2H_5NH_3^+$+$\rm C_2H_8N_3^+$, $\rm CH_3COO^-$+$\rm C_2H_8N_3^+$, $\rm CH_3COO^-$+4-methylphenole and $\rm CH_3COO^-$+4-methylimidazole. To find the MLIPs' optimal functional forms, we benchmark different MLIPs on training sets for two of the dimers, CH$_3$COO$^{-}$+CH$_3$COO$^{-}$, and CH$_3$COO$^{-}$+CH$_3$NH$_3^+$. We then analyze the binding curves for the other four dimers using the optimal MLIPs that were found using the above procedure.

\section{Methodology}

\subsection{Interatomic potentials}

In this subsection, we describe the interatomic interaction models used in this work. We first introduce the short-range models, namely, the Moment Tensor Potential (MTP) (Subsection~\ref{section:method_mtp}) and the Equivariant Tensor Network (ETN) potential (Subsection~\ref{section:method_etn}). Next, we present the long-range model developed in this study and its combination with the short-range models~(Subsection~\ref{section:method_qrd}). 

\subsubsection{Moment Tensor Potential}
\label{section:method_mtp}

MTP~\cite{MTPmulti} is a local interatomic potential for which the energy $E^{\rm short}$ of a configuration $\bm x$ is given as a sum of contributions $V_i$ of each of the $N$ atoms of the configuration:
\begin{equation}
\label{eq:en_local_pot}
    E^{\rm short} = E^{\rm }\left(\bm x\right) = \sum_{i=1}^N V_i = \sum_{i=1}^N V({\bf \mathfrak{n}}_i)
    .
\end{equation}
Here, ${\bf \mathfrak{n}}_i$ is the neighborhood of the $i$-th atom in the configuration, which is the set of all atoms in the system that are within a distance less than the cutoff radius $R_\text{cut}$ from the $i$-th atom. 

In MTP, $V_i$ is a linear combination of basis functions $B_{\alpha}$:
\begin{equation}
\label{eq:en_mtp}
    V({\bf \mathfrak{n}}_i) \equiv V^{\rm MTP}({\bf \mathfrak{n}}_i) = \sum_\alpha \xi_\alpha B_\alpha({\bf \mathfrak{n}}_i),
\end{equation}
where $\xi_\alpha$ are the linear parameters to be fitted. To construct the basis functions $B_\alpha$ we introduce the so-called moment tensor descriptors:
\begin{equation}
\label{eq:moment_tensor_descriptor}
    M_{\mu, \nu}({\bf \mathfrak{n}}_i) = \sum_j f_\mu\left(r_{ij}, z_i, z_j\right)\bm r_{ij}^{\otimes \nu},
\end{equation}
where $\bm r_{ij} = \bm r_{j} - \bm r_{i}$, $r_{ij}=\left|\bm r_{ij}\right|$ are the relative positions and distances between atoms $i$ and $j$, respectively, and $z_i$ and $z_j$ are the types of these atoms. The symbol ``$\otimes$'' denotes the outer product of vectors and, thus, the second part in~\eqref{eq:moment_tensor_descriptor}, $\mathbf{r}_{ij}^{\otimes \nu}$, is an angular part which represents a tensor of order $\nu$. The first part in~\eqref{eq:moment_tensor_descriptor}, $f_\mu$, is a radial part, which is given as a linear combination of radial functions $\varphi^{\left(\beta\right)}$:
\begin{equation}
\label{eq:mtp_radial}
    f_\mu\left(r_{ij}, z_i, z_j\right) = \sum_\beta c_{\mu, z_i, z_j}^{\left(\beta\right)}\varphi^{\left(\beta\right)}\left(r_{ij}\right)
    ,
\end{equation}
where $c_{\mu, z_i, z_j}^{\left(\beta\right)}$ is another set of parameters to be fitted.
We note that each of the $\varphi^{\left(\beta\right)}$'s tends to zero as $r_{ij}$ tends to $R_\text{cut}$. In this work, radial basis set is comprised of the following infinitely differentiable functions:
\begin{equation}
    \label{eq:hat_func}
    \varphi^{(\beta)}(r_{ij}) = \begin{cases}
        r_{ij}^{2\beta} \textit{}e^{-\frac{1}{R_{\rm cut}^2-r_{ij}^2}}, \; &r_{ij} \le R_{\rm cut}\\
        0, &\rm otherwise.
    \end{cases}
\end{equation}
We further apply the Gram-Schmidt orthogonalization to this set of functions.

The set of the basis functions $B_{\alpha}$ is then obtained from all possible contractions of moment tensor descriptors yielding a scalar. To choose some finite subset of basis functions, the level ($\text{lev}$) of moment tensor descriptors and basis functions is introduced:
\begin{gather*}
    \text{lev} M_{\mu, \nu} = 2 + 4\mu + \nu, \\
    \text{lev} \prod_k M_{\mu_k, \nu_k} = \sum_k \text{lev} M_{\mu_k, \nu_k} = \sum_k \left(2 + 4\mu_k + \nu_k\right).
\end{gather*}
We say that an MTP is of level $\rm lev_{\rm max}$ if it includes all basis functions with the level not greater than $\rm lev_{\rm max}$. 

\subsubsection{Equivariant Tensor Network potential}
\label{section:method_etn}
The other short-range model used in this work is ETN \cite{ETN}. In this case, $V(\bm n_i)$ in \eqref{eq:en_local_pot} is a tensor network represented as the tensor train (TT) \cite{TT}:

\begin{widetext}
\begin{equation}\label{eq:etn_pot}
\begin{array}{c}
    \displaystyle
    V({\bf \mathfrak{n}}_i) \equiv V^{\rm ETN}({\bf \mathfrak{n}}_i)
    =
    \sum_{k_1} \sum_{k_2} \ldots \sum_{k_{d-2}} \sum_{k_{d-1}} \left(\sum_{k_1'} T^1_{k_1' k_1} v_{k_1'}({\bf \mathfrak{n}}_i)\right) 
    \left(\sum_{k_2'} T^2_{k_1 k_2' k_2} v_{k_2'}({\bf \mathfrak{n}}_i)\right)
    \ldots \\
    \displaystyle
    \left(\sum_{k_{d-1}'} T^{d-1}_{k_{d-2} k_{d-1}' k_{d-1}} v_{k_{d-1}'}({\bf \mathfrak{n}}_i)\right) \left(\sum_{k_d'} T^d_{k_{d-1} k_d'} v_{k_d'}({\bf \mathfrak{n}}_i)\right),
\end{array}
\end{equation}
\end{widetext}
where $T^1_{k_1' k_1}, T^2_{k_1 k_2' k_2}, \ldots, T^{d-1}_{k_{d-2} k_{d-1}' k_{d-1}}, T^d_{k_{d-1} k_d'}$ are the so-called TT cores, and $v_{k_j'}({\bf \mathfrak{n}}_i)$, $j=1,\ldots,d$ are the so-called feature vectors depending on atomic neighborhoods ${\bf \mathfrak{n}}_i$. We discuss each of these parts below.

ETN is a local interatomic interaction potential and, therefore, $V^{\rm ETN}({\bf \mathfrak{n}}_i)$ should be invariant under actions of SO(3) (group of rotations). A sufficient condition to make $V^{\rm ETN}({\bf \mathfrak{n}}_i)$ invariant to rotations is to let the TT cores be equivariant mappings of the feature vectors to rotate them correspondingly with a basis change under SO(3). To that end, we consider a decomposition of each $v_k$ into an irreducible representation of SO(3) using spherical harmonics:
\begin{equation}\label{eq:etn_feature}
    v_{k} \equiv v_{\ell mn}
    =
    \sum_j f_n(r_{ij},z_i,z_j) Y_{\ell m}(\bm r_{ij} / r_{ij}),
\end{equation}
where a multi-index $k=(\ell mn)$ is composed of $\ell = 0, \ldots, L$, the index of the subspace of the irreducible representation, $m = -\ell, -\ell + 1, \ldots, \ell$, the dimension of the subspace, and $n = 1, \ldots, N(\ell)$, the number of radial channels corresponding to each $\ell$. As in MTP, $f_n(r_{ij},z_i,z_j)$ is a radial part which is also a tensor network:
\begin{equation}\label{eq:etn_radial_basis}
    f_n(r_{ij},z_i,z_j) = \sum_{\beta\lambda} B_{n \beta \lambda} \varphi^{(\beta)}(r_{ij}) A_{\lambda z_i z_j}
    ,
\end{equation}
in which the tensors $A$ and $B$ are parameters to be fitted (see details in Ref.~\onlinecite{ETN}), and $Y_{\ell m}(\bm r_{ij} / r_{ij})$ are spherical harmonics containing angular contributions.

To make the TT cores equivariant, we employ the Wigner-Eckhart theorem stating that every tensor with three multi-indices $\{ (\ell_i, m_i, n_i) \}_{i=1,\ldots,3}$ can be represented in the following form:
\begin{equation}\label{eq:etn_core}
    T_{(\ell_1 m_1 n_1)(\ell_2 m_2 n_2)(\ell_3 m_3 n_3)}
    =
    \xi_{(\ell_1 n_1)(\ell_2 n_2)(\ell_3 n_3)}
    C_{(\ell_1 m_1)(\ell_2 m_2)(\ell_3 m_3)}
    ,
\end{equation}
where $\xi_{(\ell_1 n_1)(\ell_2 n_2)(\ell_3 n_3)}$ is the tensor of coefficients to be fitted, and $C_{(\ell_1 m_1)(\ell_2 m_2)(\ell_3 m_3)}$ is the Clebsch-Gordan coefficient that defines the symmetry group. We remark that the equation \eqref{eq:etn_core} is also valid for the case of two multi-indices, i.e., for the first and the last cores in \eqref{eq:etn_pot} with $\ell_3=m_3=n_3=0$. Thus, we use two techniques to construct an ETN: the TT-representation which enables generating a tensor of any order using the tensors of the third order and the Wigner-Eckhart theorem which gives us an opportunity to represent any of the third-order tensors as equivariant mappings.


\subsubsection{Charge Redistribution}
\label{section:method_qrd}

Charge Redistribution (QRd) is the long-range model used this work. In this model, the energy of interatomic interaction is represented as the Coulomb energy of point charges centered on atoms, given by:
\begin{equation}
\label{eq:qrd_en}
    E^{\rm long} = \sum_{i<j}\frac{q_iq_j}{r_{ij}},
\end{equation}
where $q_i$ and $q_j$ are the charges of the $i$-th and $j$-th atoms, respectively. The charge of an atom is predicted using the following equation:
\begin{equation}
\label{eq:qrd_charges}
    \displaystyle
    q_i({\bm z}, {\bm a}) = q_i({\bm z}, ({\bm b}, {\bm s})) = b_{z_i} + s_{z_i} \frac{Q_{\rm total} - \sum_jb_{z_j}}{\sum_js_{z_j}},
\end{equation}
where ${\bm b}$, ${\bm c}$ are the vectors of the model parameters (overall, there are two parameters per atom type), and $Q_{\rm total}$ is the total charge of the system. We note that the sum of all charges in the system (i.e. its total charge) is always equal to $Q_{\rm total}$, which enables total charge conservation.

In this charge prediction scheme, atomic charges depend solely on the chemical composition of the system~(i.e., types of all the atoms comprising the system), and are independent of atomic positions. It immediately leads to atoms of the same type having the same charge that does not change as long as the chemical composition of the system does not change (e.g., during molecular dynamics simulations in NVE, NVT, and NpT ensembles).

We note that it is possible to modify~\eqref{eq:qrd_charges} (e.g., by making $\bm b$ environment-dependent) to obtain a model with both environment-dependent charges and total charge conservation, but we leave it for future work.

Combined models~(MTP+QRd and ETN+QRd) are introduced as a simple combination of the short-range term given by MTP or ETN and the long-range term given by QRd:
\begin{equation}
    \label{eq:combined_model_energy}
    E^{\rm total}\left(\bm x, \bm \theta, \bm a\right) = 
    E^{\rm short}\left(\bm x, \bm \theta \right) +
    E^{\rm long}\left(\bm x,  \bm a\right),
\end{equation}
where $\bm \theta$ and $\bm a$ are the parameters of the short-range and long-range parts, respectively.

\subsection{Fitting}

Assume we have $K$ configurations in the training set, and for each configuration $\bm x_{(k)}, k=1,\ldots,K$ we have energies $E^{\rm DFT}\left(\bm x_{(k)}\right)$ and forces $F^{\rm DFT}_{i, l}\left(\bm x_{(k)}\right)$ acting on the $i$-th atom ($l=1,2,3$ is the component of force), obtained using DFT calculations. To find the parameters of the short-range potential, we minimize the loss function~\eqref{eq:lossfunction_sr} with respect to the parameters of this potential:
\begin{equation}
    \label{eq:lossfunction_sr}
    \begin{split}
    \mathcal{L}_1(\bm \theta) = \sum_{k=1}^K &\left[w_e \left(E^{\rm short}\left(\bm x_{(k)}, \bm \theta\right)-E^{\rm DFT}\left(\bm x_{(k)}\right)\right)^2\right.\\
    &+\left.w_f \sum_{i=1}^N \sum_{l=1}^3 \left(F^{\rm short}_{i, l}\left(\bm x_{(k)}, \bm \theta\right)-F^{\rm DFT}_{i, l}\left(\bm x_{(k)}\right)\right)^2\right],
    \end{split}
\end{equation}
where $w_e$ and $w_f$ are non-negative weights, determining the importance of energies and forces with respect to each other.

To train a combined (long-range) model, we can utilize the fact that our model predicts not only energies, and forces, but also charges. Therefore, we use partial atomic charges $\left(q^{\rm DFT}_m\left(\bm x_{(k)}\right)\right)$ obtained with DFT, and include an additional term in the loss function~\eqref{eq:lossfunction_lr}:

\begin{widetext}
\begin{equation}
		\label{eq:lossfunction_lr}
		\begin{split}
            \mathcal{L}_2(\bm \theta, \bm a) = \sum_{k=1}^K &\left[w_q\sum_{i=1}^N \left(q_i\left(\bm x_{(k)}, \bm a\right)-q_i^{\rm DFT}\left(\bm x_{(k)}\right)\right)^2\right.\\
            &+\left.w_e \left(E^{\text{total}}\left(\bm x_{(k)}, \bm \theta,\bm a\right)-E^{\rm DFT}\left(\bm x_{(k)}\right)\right)^2+
            w_f \sum_{i=1}^N  \sum_{l=1}^3 \left(F^{\text{total}}_{i, l}\left(\bm x_{(k)}, \bm \theta,\bm a\right)-F^{\rm DFT}_{i, l}\left(\bm x_{(k)}\right)\right)^2
    \right],
			\end{split}
\end{equation}
\end{widetext}
where $w_q$ is a non-negative weight which expresses the importance of charges. Thus, to train a long-range model, we minimize $\mathcal{L}_2(\bm \theta, \bm a)$ with respect to the parameters of both short-range and long-range parts of a model simultaneously. 

Minimization of both loss functions $\mathcal{L}_1(\bm \theta)$ and $\mathcal{L}_2(\bm \theta, \bm a)$ was performed using the Broyden--Fletcher--Goldfarb--Shanno (BFGS) algorithm.

\subsection{Optimizing the ETN hyperparameters} \label{ETN_hyper_opt}

To obtain the optimal set of ETN hyperparameters for a given training set, we utilize the hyperparameter optimization procedure from Ref.~\onlinecite{ETN}.
The ETN from Ref.~\onlinecite{ETN} has in total six hyperparameters, namely,
\begin{itemize}
    \item
    the number of equivariant TT cores $d$,
    \item
    the number of angular subspaces $L$,
    \item
    the number of radial basis functions,
    \item
    the rank of the TT cores (we assume constant ranks over the entire tensor train),
    \item
    the rank of $A$ \eqref{eq:etn_radial_basis},
    \item
    and the rank of $B$ \eqref{eq:etn_radial_basis}.
\end{itemize}
The loss functions in \eqref{eq:lossfunction_sr}, \eqref{eq:lossfunction_lr} implicitly depend on the hyperparameters of the potential and on the training set $\mathcal{T}$. In this subsection, we use a modified notation to show this dependence explicitly:
\begin{equation*}
    L({\bm \theta}) \equiv L(\mathcal{H}, {\bm \theta}, \mathcal{T})
    .
\end{equation*}
Our hyperparameter optimization is as follows:
\begin{enumerate}
    \item[0)] We select the initial vector of hyperparameters $\mathcal{H}$.
    \item[1)] We train an ensemble of $N_{\rm ens}$ potentials on the training set $\mathcal{T}$ starting from random parameters and compute mean validation loss $\mathcal{L}$ on the validation set $\mathcal{V}$:
    \begin{equation}
        \mathcal{L}(\mathcal{H}, \mathcal{V}) = \frac{1}{N_{\rm ens}}\sum_{i=1}^{N_{\rm ens}}L(\mathcal{H}, {\bm \theta_i^*}, \mathcal{V}).
    \end{equation}
    \item[2)] We try to increase each hyperparameter by one individually and denote these updates by $\mathcal{H}+\Delta_i\mathcal{H}$. For each of the new potentials, we train an ensemble and calculate the mean validation loss.
    \item[3)] For each of the new potentials, we calculate the stochastic gradient of the mean validation loss function ($\nabla_i\mathcal{L}$) as follows:
    \begin{equation}
    \label{eq:stochastic_grad}
    \nabla_i\mathcal{L}(\mathcal{H}, \mathcal{V}) = - \frac{\ln\mathcal{L}(\mathcal{H}+\Delta_i\mathcal{H}, \mathcal{V})-\ln\mathcal{L}(\mathcal{H}, \mathcal{V})}{\ln\#(\mathcal{H}+\Delta_i\mathcal{H})-\ln\#(\mathcal{H})},
    \end{equation}
    where $\#(\mathcal{H})$ is an amount of parameters in the ETN with the hyperparameters $\mathcal{H}$.
    \item[4)] We choose $i$ that maximizes the stochastic gradient and increase the corresponding hyperparameter by one. Then we return to step 2).
\end{enumerate}
This procedure is repeated until either all the stochastic gradients are negative or chosen limits of optimization iterations and the number of potential parameters are exceeded.

Essentially the same procedure is used for the ETN+QRd models with two minor adjustments:
\begin{itemize}
    \item For each set of ETN hyperparameters, an ensemble of ETN+QRd's is trained instead of the ensemble of ETNs.
    \item The total number of parameters in ETN+QRd is used in \eqref{eq:stochastic_grad}.
\end{itemize}

\subsection{Investigated systems}

\begin{figure*}[!ht]
    \centering
    \begin{subfigure}[t]{0.3\linewidth}
        \centering
        \includegraphics[width=\textwidth]{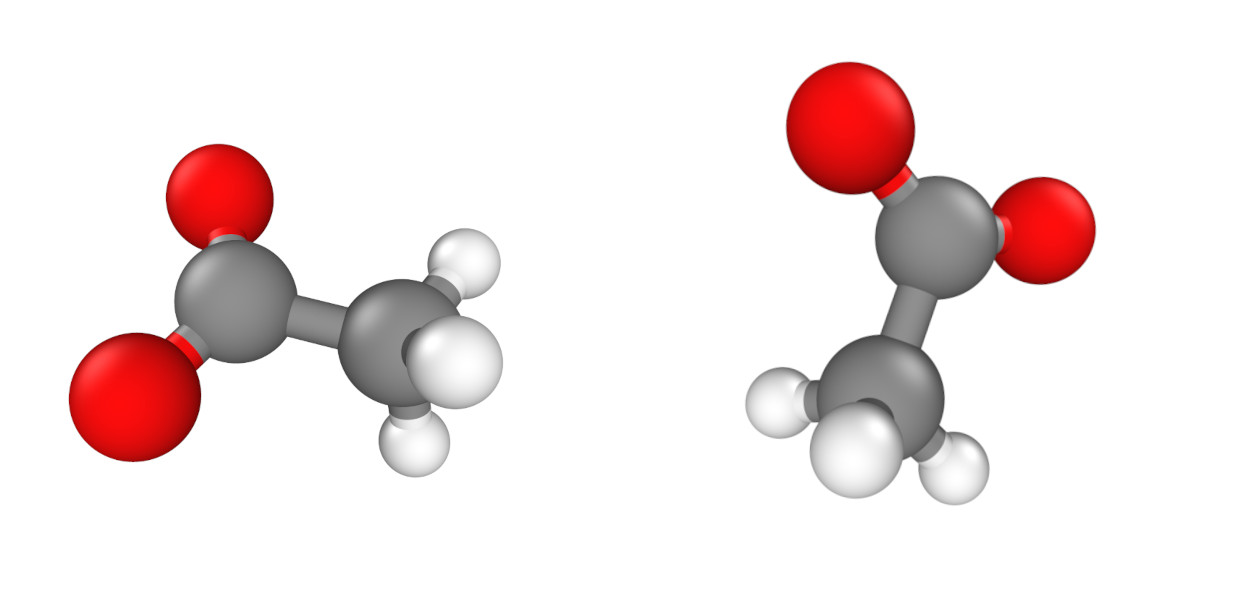}
        \caption{$\rm CH_3COO^-$+$\rm CH_3COO^-$.}
        \label{fig:system_ch3coo_2}
    \end{subfigure}\hfill
    \begin{subfigure}[t]{0.3\linewidth}
        \centering
        \includegraphics[width=\textwidth]{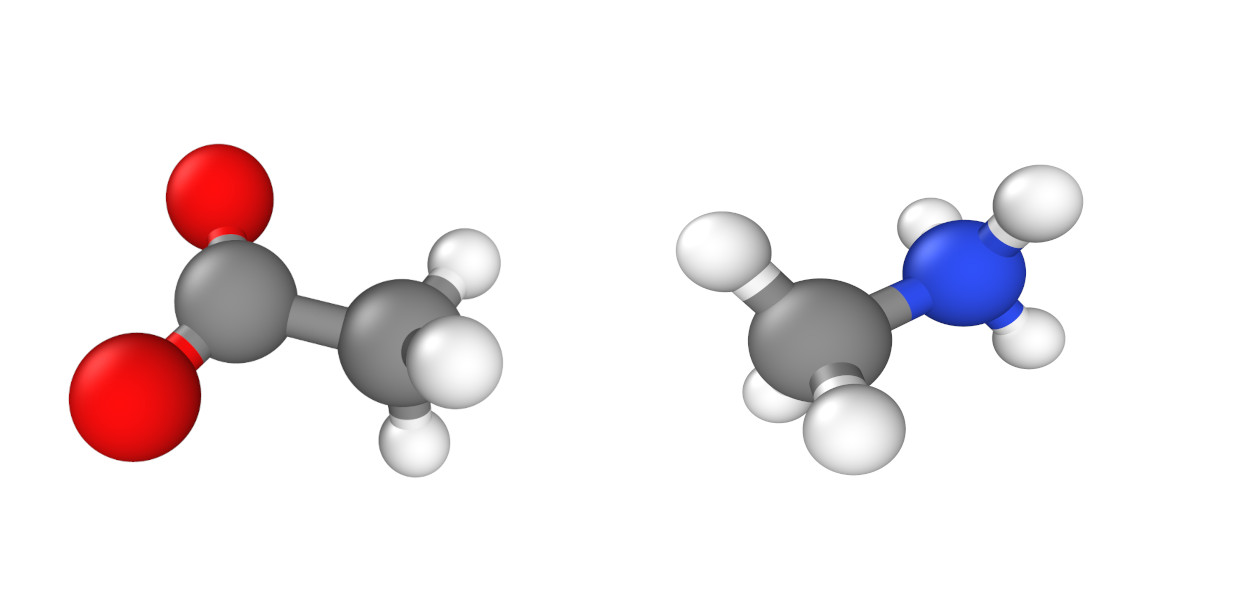}
        \caption{$\rm CH_3COO^-$+$\rm CH_3NH_3^+$.}
        \label{fig:system_ch3nh3_ch3coo}
    \end{subfigure}\hfill
    \begin{subfigure}[t]{0.3\linewidth}
        \centering
        \includegraphics[width=\textwidth]{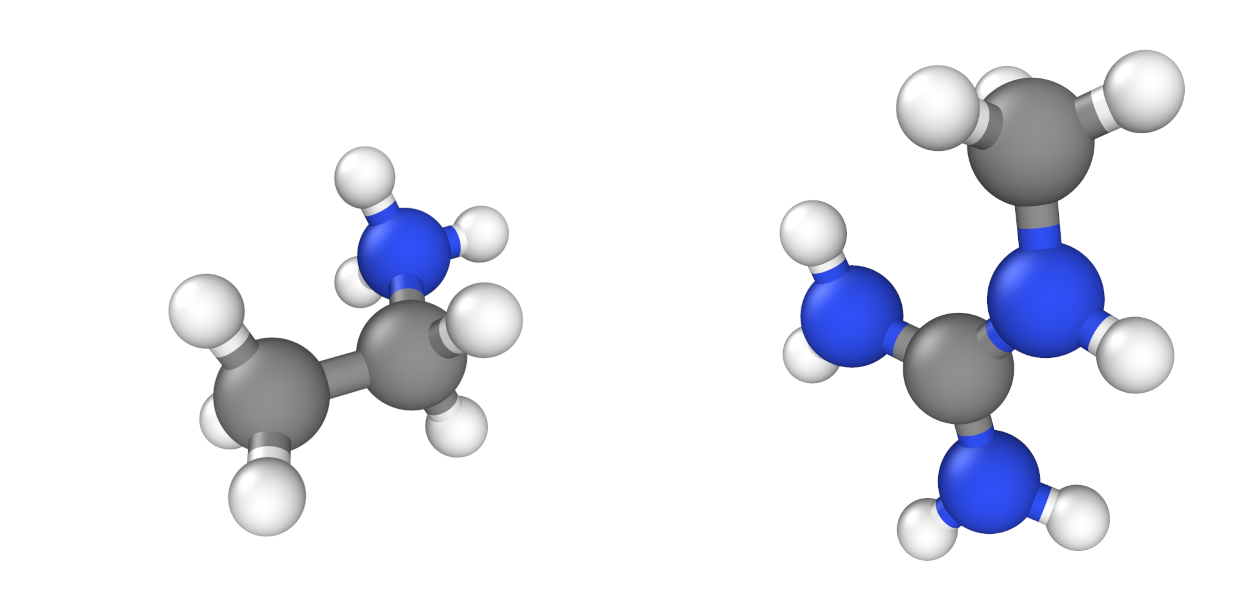}
        \caption{$\rm C_2H_5NH_3^+$+$\rm C_2H_8N_3^+$.}
        \label{fig:system_ethylammonium_c2h8n3}
    \end{subfigure}

    \vspace{0.5cm}

    \begin{subfigure}[t]{0.3\linewidth}
        \centering
        \includegraphics[width=\textwidth]{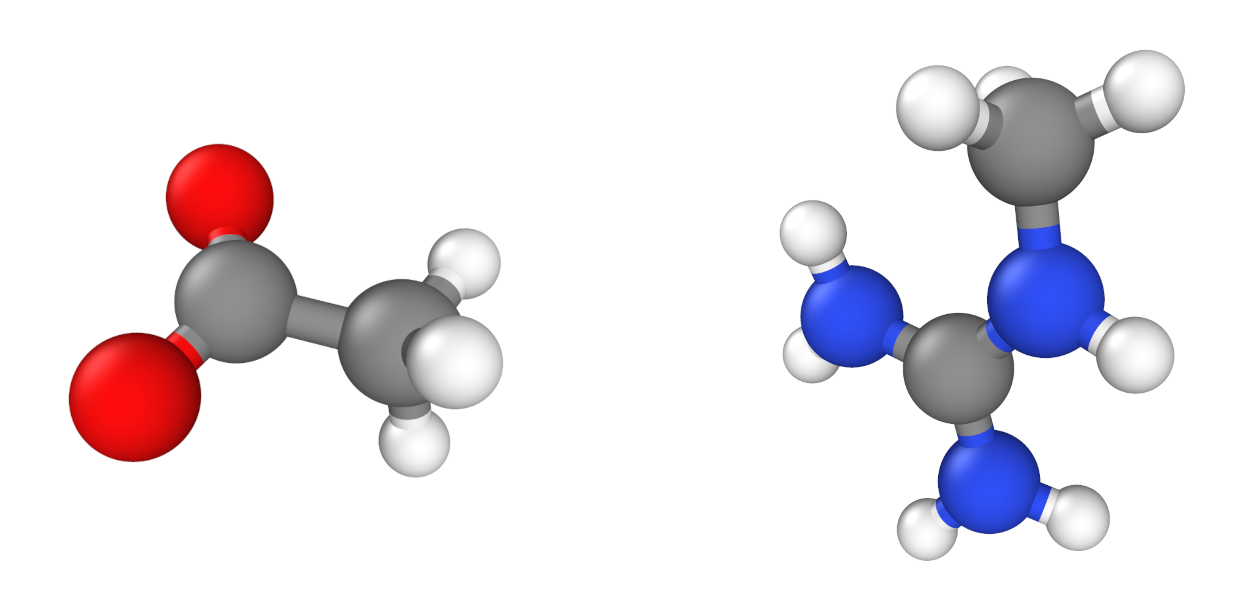}
        \caption{$\rm CH_3COO^-$+$\rm C_2H_8N_3^+$.}
        \label{fig:system_ch3coo_c2h8n3}
    \end{subfigure}\hfill
    \begin{subfigure}[t]{0.3\linewidth}
        \centering
        \includegraphics[width=\textwidth]{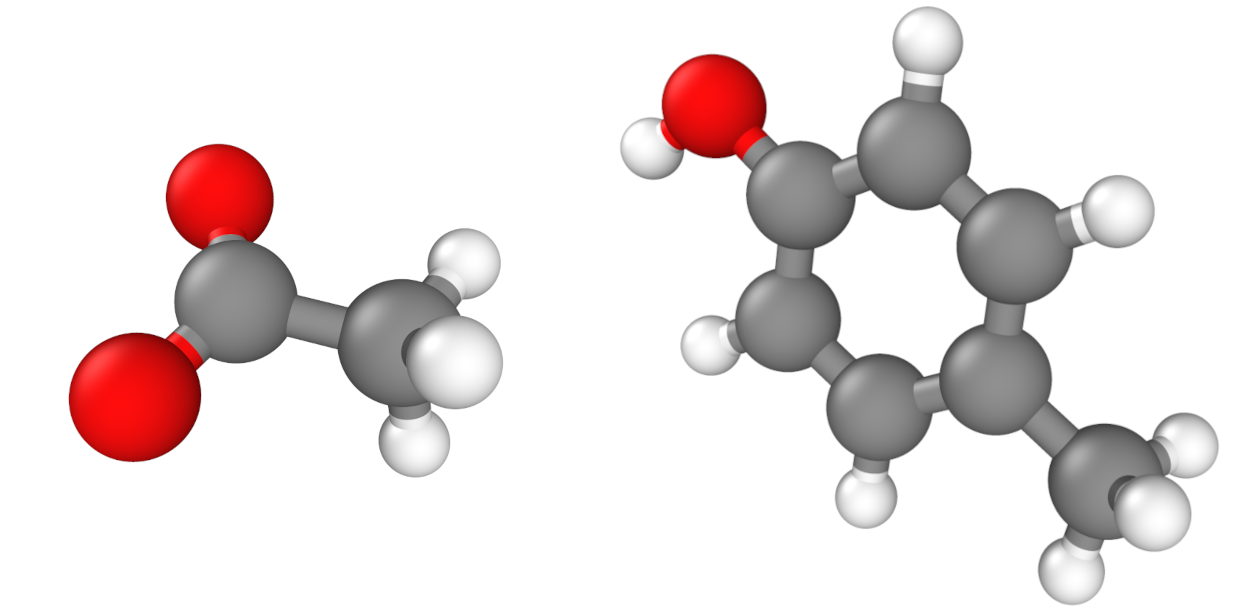}
        \caption{$\rm CH_3COO^-$+4-methylphenole.}
        \label{fig:system_ch3coo_methylphenole}
    \end{subfigure}\hfill
    \begin{subfigure}[t]{0.3\linewidth}
        \centering
        \includegraphics[width=\textwidth]{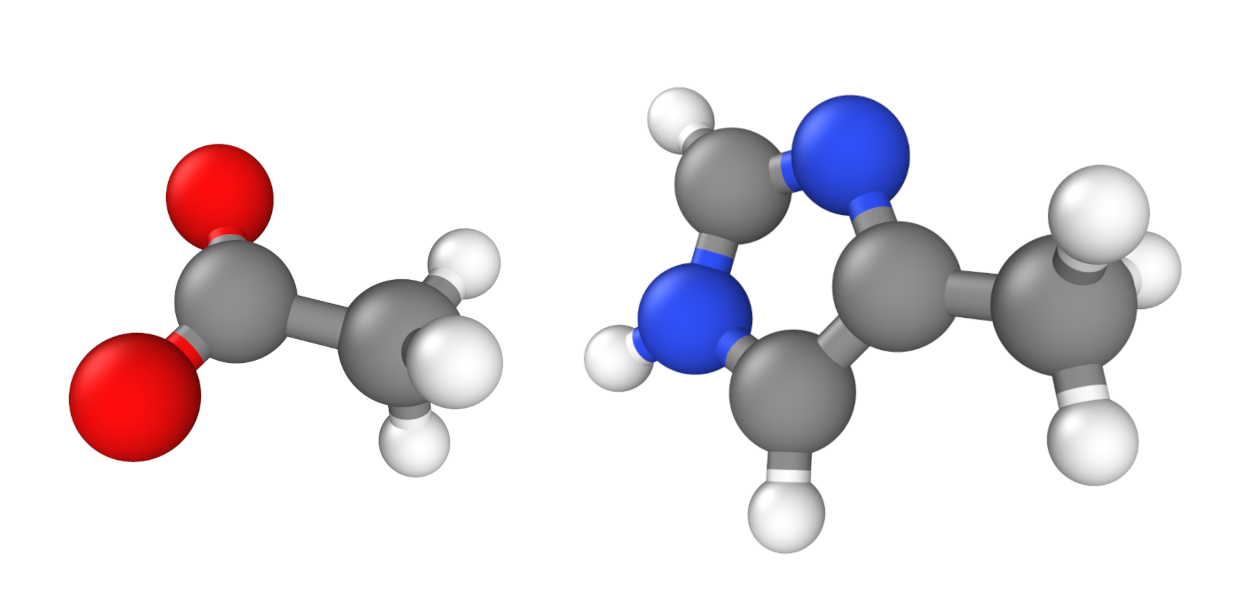}
        \caption{$\rm CH_3COO^-$+4-methylimidazole.}
        \label{fig:system_ch3coo_methylimidazole}
    \end{subfigure}
    
    \caption{Snapshots of investigated systems. Color code: white = hydrogen, grey = carbon, blue = nitrogen, red = oxygen. Visualizations were performed using OVITO~\cite{OVITO}.}
    \label{fig:binding_curves_ions}
\end{figure*}

In this work, we tested our models on six molecular dimers in vacuum: CH$_3$COO$^{-}$+CH$_3$COO$^{-}$ (Fig.~\ref{fig:system_ch3coo_2}), CH$_3$COO$^{-}$+CH$_3$NH$_3^+$ (Fig.~\ref{fig:system_ch3nh3_ch3coo}), $\rm C_2H_5NH_3^+$+$\rm C_2H_8N_3^+$ (Fig.~\ref{fig:system_ethylammonium_c2h8n3}), $\rm CH_3COO^-$+$\rm C_2H_8N_3^+$ (Fig.~\ref{fig:system_ch3coo_c2h8n3}), $\rm CH_3COO^-$+4-methylphenole (Fig.~\ref{fig:system_ch3coo_methylphenole}) and $\rm CH_3COO^-$+4-methylimidazole (Fig.~\ref{fig:system_ch3coo_methylimidazole}). The first four systems consist of dimers formed by molecular ions, where intermolecular interactions are dominated by charge-charge contribution. The last two systems are dimers composed of a molecular ion and a neutral polar molecule, with intermolecular interactions that asymptotically approach charge-dipole behavior.

\section{Results and discussion}

\subsection{Choosing the short-range MLIP}
\label{sec:res_hyperparam_opt}

In this part of the work, we investigate two systems containing dimers of organic ions in vacuum, namely $\rm CH_3COO^- + CH_3COO^-$ and $\rm CH_3COO^- + CH_3NH_3^+$, and select the short-range model and its hyperparameters for further testing.

\subsubsection{Computational details}

We constructed separate training, validation, and test sets for each system. These sets comprised several sampled binding curves (i.e., sets of configurations with the same relative orientation of molecules, but different distances between them) with molecules separation between approximately 3~\AA \;and 18~\AA. For both of the systems, training, validation, and test sets comprised 8, 2, and 2, sampled binding curves with 50 configurations each, respectively. Orientations of molecules in these curves calculations were chosen randomly.

DFT calculations of those sets were performed with CP2K~\cite{CP2K}. We used GTH-PBE pseudopotential with DFT-D3 correction~\cite{GrimmeD3} and TZV2P-MOLOPT-PBE-GTH basis set. Multi-grid contained 4 levels with the cutoff of the finest grid equal to 1000~Ry and 60~Ry cutoff for the reference grid. Charges were partitioned from the results of DFT calculations using the Mulliken method~\cite{MullikenCharges}.

Every short-range model in this part of the work has a cutoff radius of 5~\AA \;and an infinitely differentiable radial basis as described in subsection~\ref{section:method_mtp}. Further, every MTP has 8 radial basis functions.

During training and hyperparameter optimization, we trained ensembles of 20 potentials starting from random parameters and used the following weights:
\begin{equation}
    \label{eq:weights}
    w_e = 1.0 \; \left(\text{eV}\right)^{-2}, \; w_f = 0.01 \; \left(\text{eV/\AA}\right)^{-2}, \; w_q = 1.0 \; e^{-2}.
\end{equation}
In hyperparameter optimization, we used the initial set of ETN hyperparameters introduced in Ref.~\onlinecite{ETN}. The optimization was stopped when at least one of the three following conditions was satisfied:
\begin{itemize}
    \item The maximum stochastic gradient calculated using 3 best potentials in the ensemble is negative.
    \item More than 20 steps of hyperparameter optimization were performed.
    \item The obtained potential had more than 500 parameters.
\end{itemize}

\subsubsection{Optimization of hyperparameters}

\begin{figure}[!ht]
    \centering
    \begin{subfigure}{\linewidth}
        \centering
        \includegraphics[width=1.0\textwidth]{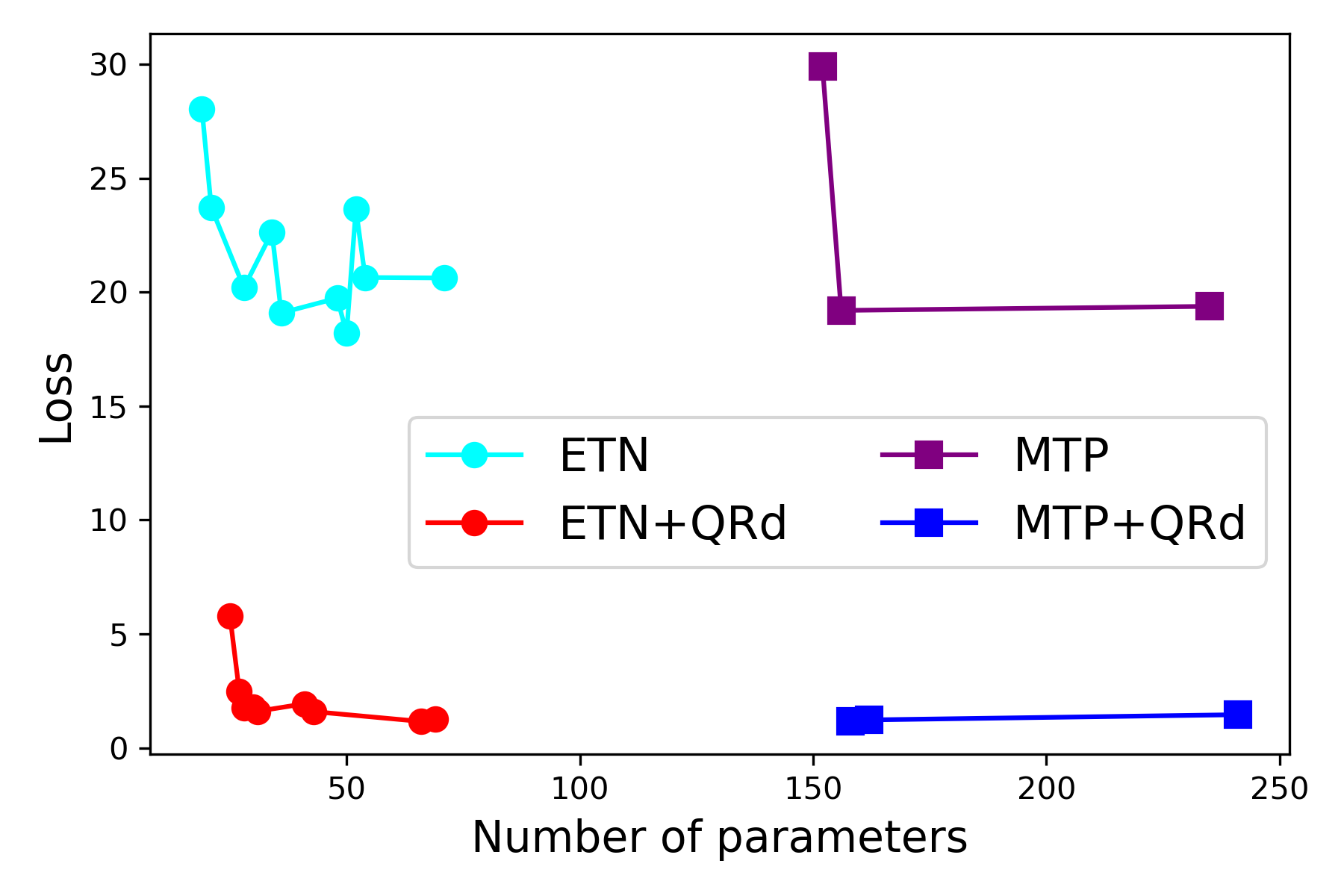}
        \caption{CH$_3$COO$^{-}$+CH$_3$COO$^{-}$.}
        \label{fig:ch3coo_hyper_optimization}
    \end{subfigure}
    
    \vspace{0.5cm} 
    
    \begin{subfigure}{\linewidth}
        \centering
        \includegraphics[width=1.0\textwidth]{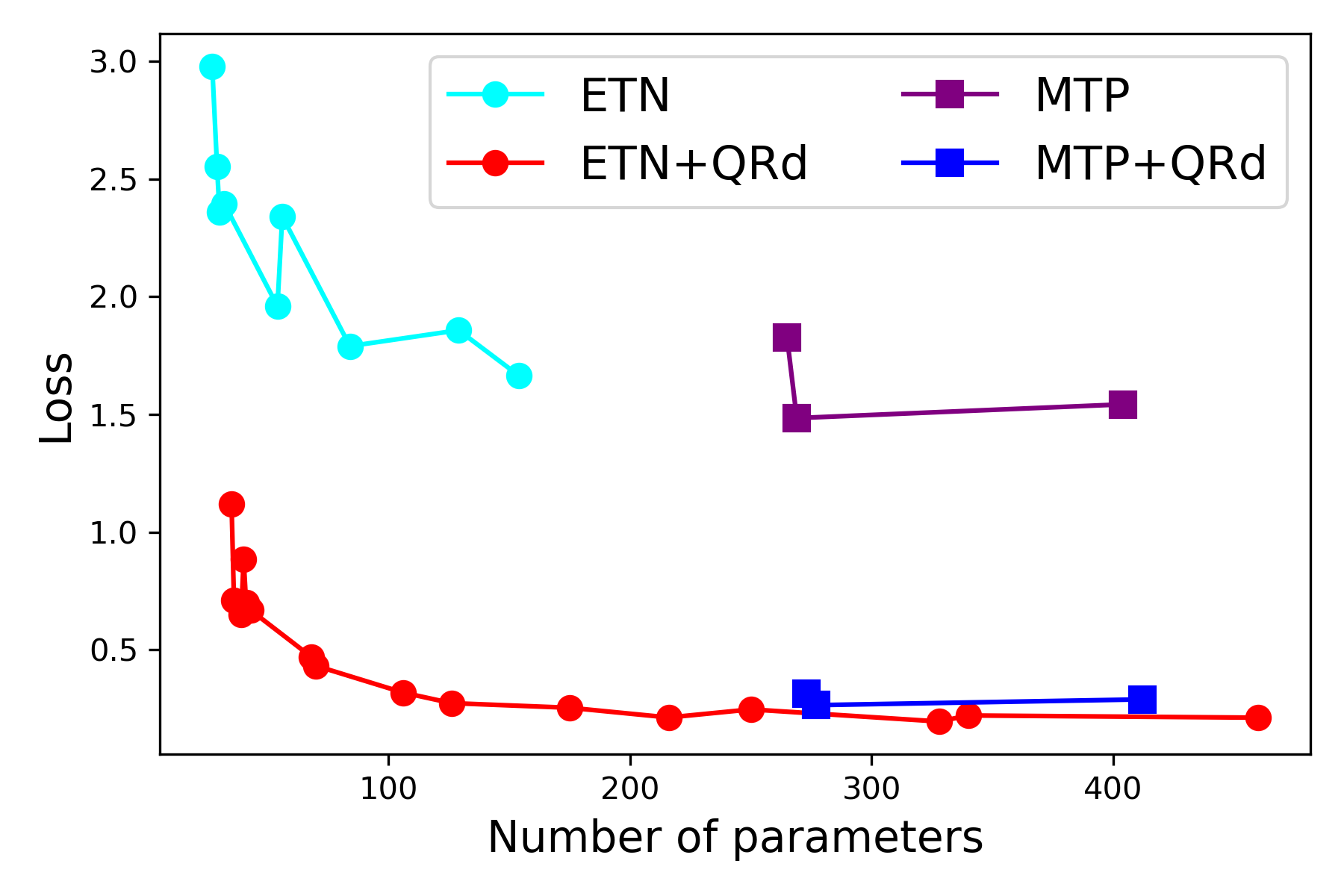}
        \caption{CH$_3$COO$^{-}$+CH$_3$NH$_3^+$.}
        \label{fig:ch3nh3hyper_optimization}
    \end{subfigure}

    \caption{Dependence of the loss function on the number of parameters of the considered MLIPs for the CH$_3$COO$^{-}$+CH$_3$COO$^{-}$ (a) and CH$_3$COO$^{-}$+CH$_3$NH$_3^+$ (b) systems.}
    \label{fig:hyper_optimization}
\end{figure}

For each type of model, we started from the optimization of hyperparameters. For the ETN-based models, we used the technique described in subsection \ref{ETN_hyper_opt}. For the MTP-based models, we tested $\rm lev_{\rm max}$ of 6, 8, and 10. For both the ETN-based and the MTP-based models, we chose the optimal ones based on the ratio of the number of parameters and the value of the loss function. The dependence of the loss function on the number of parameters of MLIPs is shown in Fig.~\ref{fig:hyper_optimization} for the CH$_3$COO$^{-}$+CH$_3$COO$^{-}$ and CH$_3$COO$^{-}$+CH$_3$NH$_3^+$ dimers. We note that we calculated the loss function on the test set using the formula \eqref{eq:lossfunction_sr} for the short-range models and using \eqref{eq:lossfunction_lr} with $w_q=0$ for the long-range models as we compared the values of the loss functions for all four MLIPs and the functional forms of the loss functions should be similar to be able to compare them. Furthermore, for any number of parameters in both MTP+QRd and ETN+QRd we obtained the same charge root-mean square error (RMSE) of 0.166 e for CH$_3$COO$^{-}$+CH$_3$COO$^{-}$ and of 0.14 e for CH$_3$COO$^{-}$+CH$_3$NH$_3^+$. For each number of parameters, we provide the smallest loss functions from the ensemble of the fitted potentials, i.e. we demonstrate the results for the ``best'' MLIPs from the ensembles. From the figure we see that for all the MLIPs and both of the dimers we observe the convergence of the loss function in the number of parameters. From Fig. \ref{fig:hyper_optimization} we conclude that the models explicitly including charges significantly decrease the loss functions obtained with the short-range models. We also emphasize that the ETN-based models required less parameters than the MTP-based models to reach the same accuracy, which corresponds to the findings in Ref.~\onlinecite{ETN}. The chosen number of parameters for each model is given in Table \ref{Table:num_parameters}. We use the models with these numbers of parameters for further investigations. We note that all the MTP-based models in Table~\ref{Table:num_parameters} are of $\rm lev_{\rm max}=8$.

\begin{table}[!ht]
\caption{Optimal number of parameters for MTP, ETN, MTP+QRd, and ETN+QRd.}
\label{Table:num_parameters}
\begin{center}
\begin{tabular}{c|c|c}
\hline
\hline
\multirow{2}{*}{Model} & \multicolumn{2}{c}{\# parameters} \\ \cline{2-3}
& CH$_3$COO$^{-}$+CH$_3$COO$^{-}$ &  CH$_3$COO$^{-}$+CH$_3$NH$_3^{+}$ \\ \hline
ETN & 71 & 84 \\
\hline
ETN+QRd & 63+6 & 98+8\\
\hline
MTP & 156 & 269\\
\hline
MTP+QRd & 156+6 & 269+8\\
\hline
\hline
\end{tabular}
\end{center}
\end{table}

\subsubsection{Training errors}

Fitting energy, force, and charge RMSEs for the MLIPs from Table \ref{Table:num_parameters} are given in Table \ref{Table:errors_ch3coo} (for the CH$_3$COO$^{-}$+CH$_3$COO$^{-}$ system) and in Table \ref{Table:errors_ch3nh3} (for the CH$_3$COO$^{-}$+CH$_3$NH$_3^{+}$ system). We see that both energy and force RMSEs are much smaller for the long-range potentials than for the short-range ones: adding QRd decreases energy RMSEs by a factor of 9 for the CH$_3$COO$^{-}$+CH$_3$COO$^{-}$ system and by a factor of 4.5 for the CH$_3$COO$^{-}$+CH$_3$NH$_3^{+}$ system and force RMSEs by a factor of 1.5 for both systems. Therefore, explicit long-range interactions even with fixed partial charges play an important role. Furthermore, we conclude that MTP+QRd and ETN+QRd gave similar energy RMSEs, and, thus, it does not matter which short-range model (ETN or MTP) is combined with the Coulomb model and which model to use for calculating dimer binding curves (see the next subsection). 

\begin{table}[!ht]
\caption{Training errors of the MLIPs fitted to the CH$_3$COO$^{-}$+CH$_3$COO$^{-}$ system.}
\label{Table:errors_ch3coo}
\begin{center}
\begin{tabular}{c|c|c|c}
\hline
\hline
\multirow{2}{*}{Model} & energy RMSE, & force RMSE, & charge RMSE, \\
& meV/atom & meV/\AA & e \\ \hline
\hline 
ETN & 23.7 & 143 & - \\
\hline
ETN+QRd & 2.6 & 80 & 0.162 \\
\hline
MTP & 22.7 & 135 & -\\
\hline
MTP+QRd & 2.3 & 62 & 0.162 \\
\hline
\hline
\end{tabular}
\end{center}
\end{table}

\begin{table}[!ht]
\caption{Training errors of the MLIPs fitted to the CH$_3$COO$^{-}$+CH$_3$NH$_3^{+}$ system.}
\label{Table:errors_ch3nh3}
\begin{center}
\begin{tabular}{c|c|c|c}
\hline
\hline
\multirow{2}{*}{Model} & energy RMSE, & force RMSE, & charge RMSE, \\
& meV/atom & meV/\AA & e \\ \hline
\hline 
ETN & 9.9 & 98 & - \\
\hline
ETN+QRd & 2.1 & 63 & 0.137 \\
\hline
MTP & 9.2 & 80 & -\\
\hline
MTP+QRd & 2.0 & 46 & 0.137 \\
\hline
\hline
\end{tabular}
\end{center}
\end{table}

\subsubsection{Binding curves}

\begin{figure*}[!ht]
    \centering
    \includegraphics[width=1.0\linewidth]{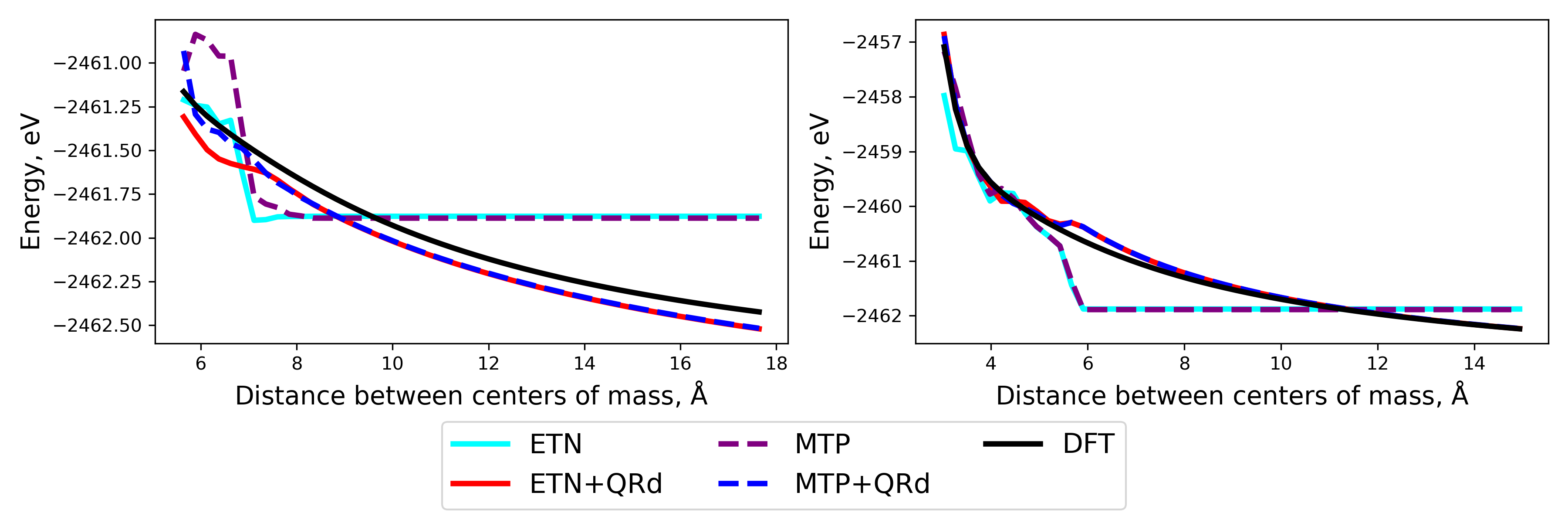}
    \caption{Dimer binding curves calculated with DFT and the MLIPs for the CH$_3$COO$^{-}$+CH$_3$COO$^{-}$ system on the test set.}
    \label{fig:binding_curves_ch3coo}
\end{figure*}

\begin{figure*}[!ht]
    \centering
    \includegraphics[width=1.0\linewidth]{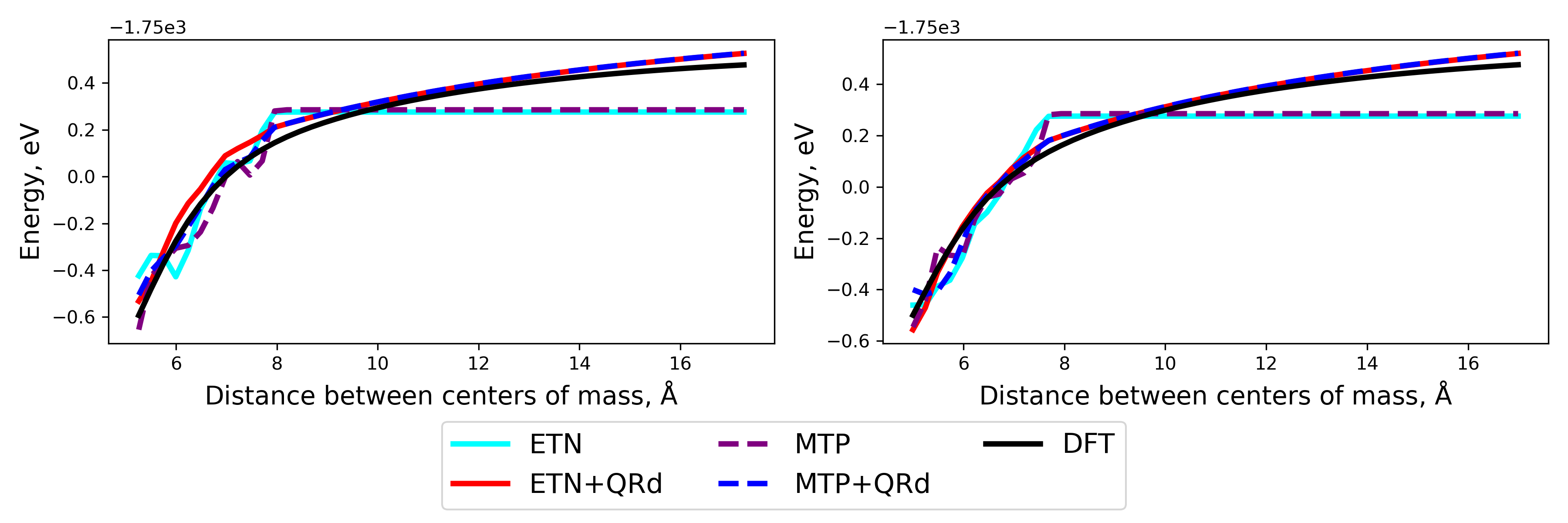}
    \caption{Dimer binding curves calculated with DFT and the MLIPs for the CH$_3$COO$^{-}$+CH$_3$NH$_3^{+}$ system on the test set.}
    \label{fig:binding_curves_ch3nh3}
\end{figure*}

Finally, we provide the binding curves calculated for the dimers from the test set. The binding curves calculated with the MLIPs and DFT are shown in Fig. \ref{fig:binding_curves_ch3coo} (for the CH$_3$COO$^{-}$+CH$_3$COO$^{-}$ system) and in Fig. \ref{fig:binding_curves_ch3nh3} (for the CH$_3$COO$^{-}$+CH$_3$NH$_3^{+}$ system). We see that the short-range MTP and ETN are overall able to capture the short-range interactions rather accurately, but they become less and less accurate as the distance between molecules increases. Starting from a point beyond the cutoff radius of short-range MLIPs, they give a constant value of energy because they predict only the separate energies of each molecule and are unable to capture long-range interactions. In contrast, the long-range MTP+QRd and ETN+QRd very accurately predict the energy of dimers' interactions for both short and long distances between the molecules. 

To quantitatively measure the improvement in the prediction of dimer binding curves with the long-range MLIPs compared to the short-range MLIPs, we calculated the $L^2$-norm of the difference of the energies predicted using DFT and the MLIPs. $L^2$-norm of a $L^2$-integrable function $f$ is defined as follows:
\begin{equation} \label{eq:binding_norm}
    \Vert f\Vert_{L^2\left(\left[a, b\right]\right)} = \sqrt{\int_a^b\left(f(x)\right)^2dx}.
\end{equation}
Here, $f$ represents the energy discrepancy $E^{\rm model} - E^{\rm DFT}$, where ``model'' is ETN, MTP, MTP+QRd, or ETN+QRd, and $\left[a, b\right]$ defines the distance range between centers of mass of molecules. These $L^2$-norms are given in Table \ref{Table:binding_curve_errors}. From the table we make two main conclusions. First, for each curve and for each system the long-range models give at least 2.5 times smaller deviations from DFT in $L^2$-norm than the short-range models. Thus, explicitly including Coulomb interactions significantly improves the accuracy of the binding curves calculations. Second, the MTP-based short-range and long-range models give results similar to the ETN-based model, and, therefore, the results obtained with the combined short-range and long-range model do not depend on the used short-range MLIP.

\begin{table*}[!ht]
\caption{$L^2$-norms of the deviations between binding energies predicted with DFT and the MLIPs.}
\label{Table:binding_curve_errors}
\begin{center}
\begin{tabular}{c|c|c|c|c}
\hline
\hline
\multirow{2}{*}{$L^2$-norm, eV$\cdot \text{\AA}^{0.5}$} & \multicolumn{2}{c|}{CH$_3$COO$^{-}$+CH$_3$COO$^{-}$} & \multicolumn{2}{c}{CH$_3$COO$^{-}$+CH$_3$NH$_3^{+}$} \\
\cline{2-5} 
& curve \# 1 & curve \# 2 & curve \# 1 & curve \# 2 \\
\hline 
$\left\Vert E^{\rm ETN} - E^{\rm DFT}\right\Vert_{L^2\left(\left[a, b\right]\right)}$ & 1.10 & 1.70 & 0.42 & 0.41 \\
\hline
$\left\Vert E^{\rm MTP} - E^{\rm DFT}\right\Vert_{L^2\left(\left[a, b\right]\right)}$ & 1.13 & 1.64 & 0.39 & 0.38 \\
\hline 
$\left\Vert E^{\rm ETN+QRd} - E^{\rm DFT}\right\Vert_{L^2\left(\left[a, b\right]\right)}$ & 0.35 & 0.33 & 0.15 & 0.09 \\
\hline
$\left\Vert E^{\rm MTP+QRd} - E^{\rm DFT}\right\Vert_{L^2\left(\left[a, b\right]\right)}$ & 0.30 & 0.31 & 0.12 & 0.12 \\
\hline
\hline
\end{tabular}
\end{center}
\end{table*}

\subsection{Additional testing}

In this section, we present the results obtained for the other four dimers in vacuum. To that end, we first describe the details of the DFT calculations, training procedure, and choice of short-range MLIPs (Subsection~\ref{sec:results_test_comp_det}). Next, we show the results for two systems containing only charged molecules: $\rm C_2H_5NH_3^+$+$\rm C_2H_8N_3^+$ and $\rm CH_3COO^-$+$\rm C_2H_8N_3^+$ (Subsection~\ref{sec:results_ions}). Finally, we show the results for two systems, which contain one charged and one neutral molecule: $\rm CH_3COO^-$+4-methylphenole and $\rm CH_3COO^-$+4-methylimidazole~(Subsection~\ref{sec:results_ions_mols}).

\subsubsection{Computational details}
\label{sec:results_test_comp_det}

As shown previously, the choice of short-range MLIP does not affect the results.
Therefore, from now on, we used only the level-8 MTP with 8 radial basis functions as a short-range model.

Since MTP does not require a validation set for its fitting, we constructed only training and test sets for each system. For all four systems, the training set contained 9 sampled binding curves, and the test set contained 1 sampled binding curve; there are approximately 35 configurations for each binding curve. Orientations of the molecules in these binding curves calculations were chosen randomly.

DFT calculations of those sets were performed using FHI-aims~\cite{FHI-AIMS}. We used PBE functional with Tkatchenko-Scheffler dispersion correction~\cite{dispersion_TS} and intermediate basis set and integration settings. Charges were partitioned from the results of DFT calculations using the Mulliken method.

The following weights of energies and forces were used for the training:
\begin{equation}
    w_e = 1.0 \; \left(\text{eV}\right)^{-2}, \; w_f = 0.01 \; \left(\text{eV/\AA}\right)^{-2}.
\end{equation}
In the following, we also investigate the effect of the weight of the charges on the training errors. Ensembles of 20 potentials with randomly initialized parameters were trained and only the ones that gave the smallest training energy errors were used for the calculation of the binding curves.

\subsubsection{Systems containing only ions}
\label{sec:results_ions}

For the $\rm C_2H_5NH_3^+$+$\rm C_2H_8N_3^+$ and $\rm CH_3COO^-$+$\rm C_2H_8N_3^+$ systems, we used an MTP with a cutoff radius of 5 \AA \;as the short-range model. For both systems, a weight of charges of $0.01 \; e^{-2}$ gave the long-range models with the lowest training errors in energies. The corresponding fitting errors for the MTP and the MTP+QRd are shown in Table~\ref{Table:errors_ions}. The binding curves for both of the systems calculated using DFT and the MLIPs are shown in Fig.~\ref{fig:binding_curves_ions}.

\begin{table*}[!ht]
\caption{Training errors for the $\rm C_2H_5NH_3^+$+$\rm C_2H_8N_3^+$ and $\rm CH_3COO^-$+$\rm C_2H_8N_3^+$ systems.}
\label{Table:errors_ions}
\begin{center}
\begin{tabular}{c|c|c|c|c}
\hline
\hline
\multirow{2}{*}{System} & \multirow{2}{*}{Model} & energy RMSE, & force RMSE, & charge RMSE, \\
& & meV/atom & meV/\AA & e \\ \hline
\multirow{2}{*}{$\rm C_2H_5NH_3^+$+$\rm C_2H_8N_3^+$} & MTP & 12.7 & 98 & -- \\ \cline{2-5}
& MTP+QRd & 1.9 & 52 & 0.142 \\ \hline 
\hline
\multirow{2}{*}{$\rm CH_3COO^-$+$\rm C_2H_8N_3^+$} & MTP & 6.2 & 89 & -- \\ \cline{2-5}
& MTP+QRd & 1.3 & 79 & 0.156 \\ \hline 
\hline
\end{tabular}
\end{center}
\end{table*}

\begin{figure}[!ht]
    \centering
    \begin{subfigure}[t]{1.0\linewidth}
        \centering
        \includegraphics[width=\textwidth]{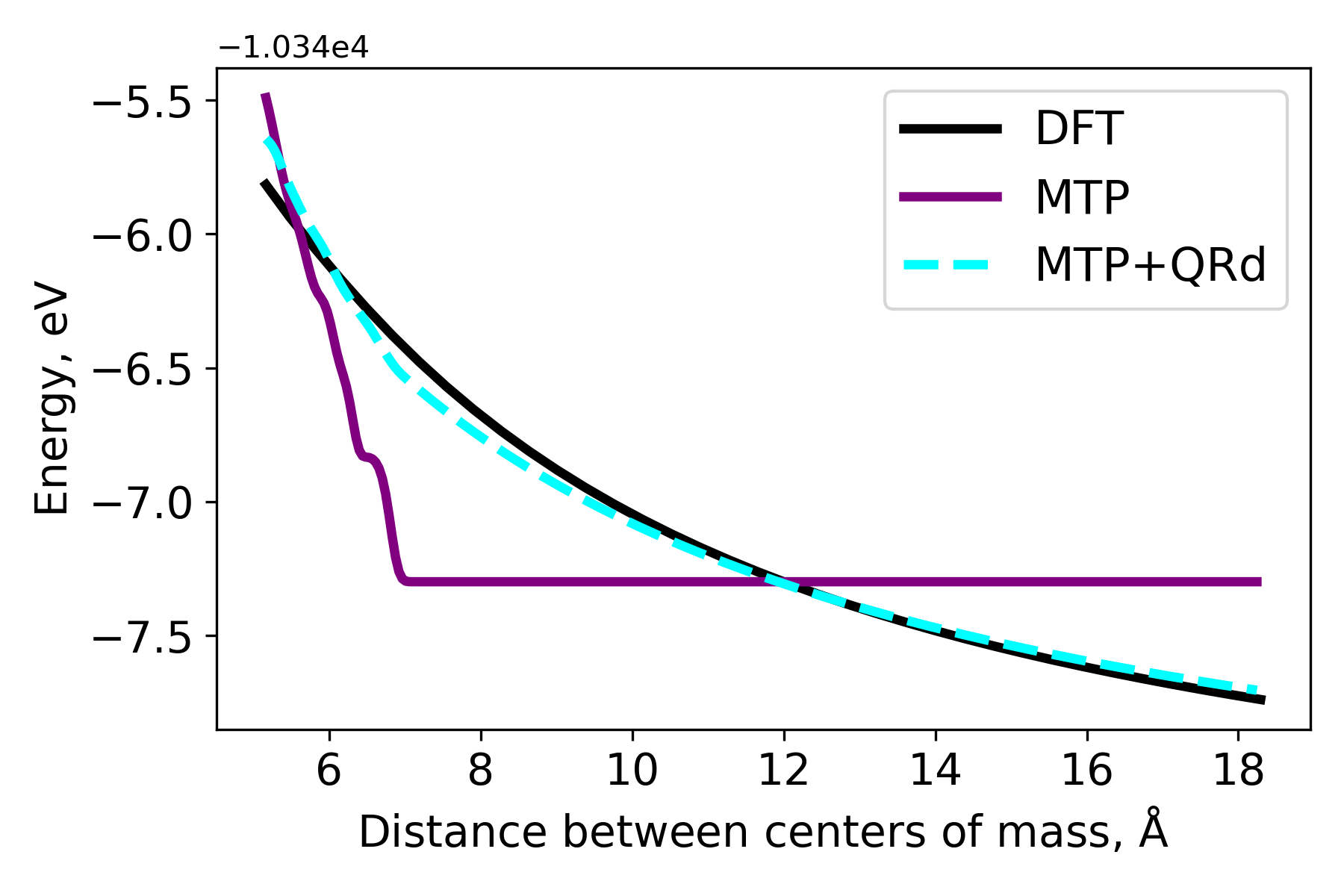}
        \caption{$\rm C_2H_5NH_3^+$+$\rm C_2H_8N_3^+$.}
        \label{fig:binding_curve_ethylammonium}
    \end{subfigure}\hfill

    \vspace{0.5cm}
    
    \begin{subfigure}[t]{1.0\linewidth}
        \centering
        \includegraphics[width=\textwidth]{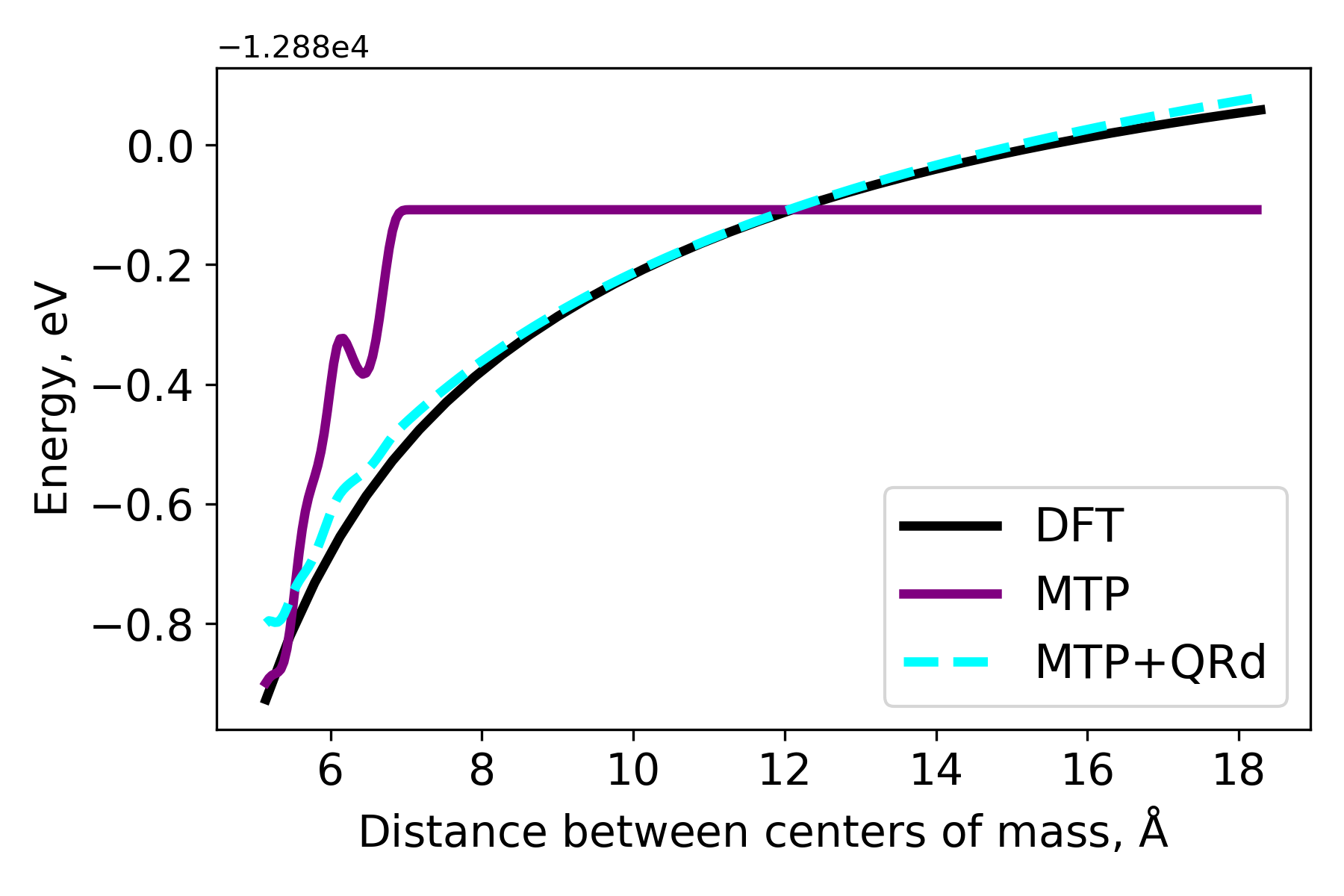}
        \caption{$\rm CH_3COO^-$+$\rm C_2H_8N_3^+$.}
        \label{fig:binding_curve_c2h8n3}
    \end{subfigure}
    \caption{Binding curves calculated using DFT and the MLIPs for the $\rm C_2H_5NH_3^+$+$\rm C_2H_8N_3^+$~(a) and $\rm CH_3COO^-$+$\rm C_2H_8N_3^+$~(b) systems.}
    \label{fig:binding_curves_ions}
\end{figure}

As above, adding QRd to the MTP, both, significantly decreases fitting energy errors, and increases the accuracy of the binding curve prediction, resulting in an excellent agreement between the binding curves obtained using DFT and MTP+QRd.

\subsubsection{Systems containing ion and neutral molecule}
\label{sec:results_ions_mols}

For the $\rm CH_3COO^-$+4-methylphenole and $\rm CH_3COO^-$+4-methylimidazole systems, we used MTPs with cutoffs of 5 and 9 \AA \;as the short-range models. Here, a weight of charges of $10^{-5} \; e^{-2}$ gave the MTP+QRd models with lowest training errors in energies. Training errors for the MTP and the MTP+QRd for both systems are shown in Table~\ref{Table:errors_ion_and_mol}. 

\begin{table*}[!ht]
\caption{Training errors for the $\rm CH_3COO^-$+4-methylphenole and $\rm CH_3COO^-$+4-methylimidazole systems.}
\label{Table:errors_ion_and_mol}
\begin{center}
\begin{tabular}{c|c|c|c|c}
\hline
\hline
\multirow{3}{*}{System} & \multirow{3}{*}{Model} & \multicolumn{3}{|c}{Training RMSEs} \\ \cline{3-5}
& & energy, & force, & charge, \\
& & meV/atom & meV/\AA & e \\ \hline
\multirow{4}{*}{$\rm CH_3COO^-$+4-methylphenole} & MTP(5\AA) & 1.0 & 38 & -- \\ \cline{2-5}
& MTP(5\AA)+QRd & 0.9 & 40 & 0.263 \\ \cline {2-5}
& MTP(9\AA) & 0.7 & 35 & -- \\ \cline{2-5}
& MTP(9\AA)+QRd & 0.6 & 41 & 0.261 \\ \hline 
\hline
\multirow{4}{*}{$\rm CH_3COO^-$+4-methylimidazole} & MTP(5\AA) & 1.8 & 45 & -- \\ \cline{2-5}
& MTP(5\AA)+QRd & 1.7 & 28 & 0.259 \\ \cline {2-5}
& MTP(9\AA) & 0.9 & 23 & -- \\ \cline{2-5}
& MTP(9\AA)+QRd & 0.9 & 27 & 0.247 \\ \hline 
\hline
\end{tabular}
\end{center}
\end{table*}

\begin{figure}[!ht]
    \centering
    \begin{subfigure}[t]{1.0\linewidth}
        \centering
        \includegraphics[width=\textwidth]{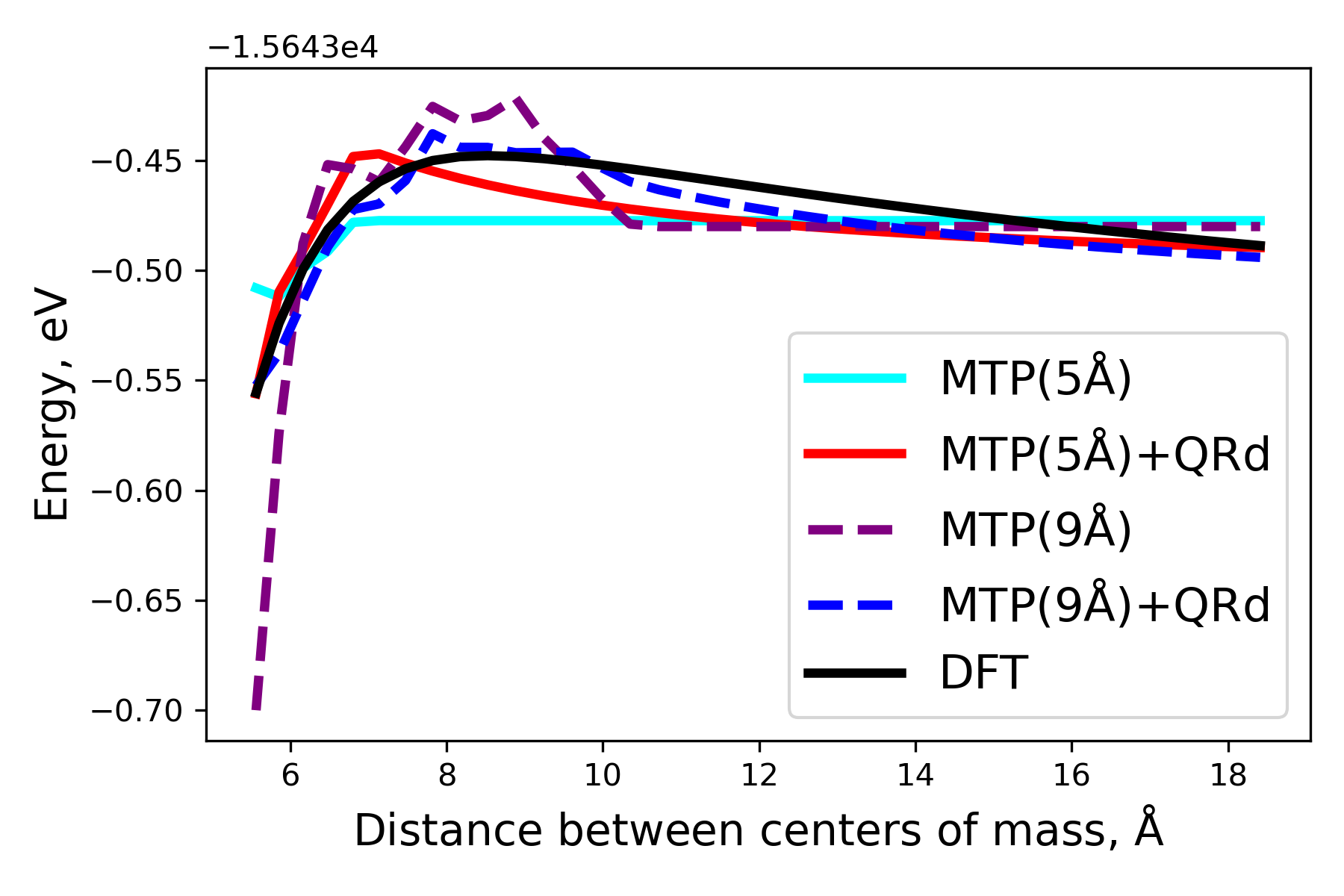}
        \caption{$\rm CH_3COO^-$+4-methylphenole.}
        \label{fig:binding_curve_phenole}
    \end{subfigure}\hfill

    \vspace{0.5cm}
    
    \begin{subfigure}[t]{1.0\linewidth}
        \centering
        \includegraphics[width=\textwidth]{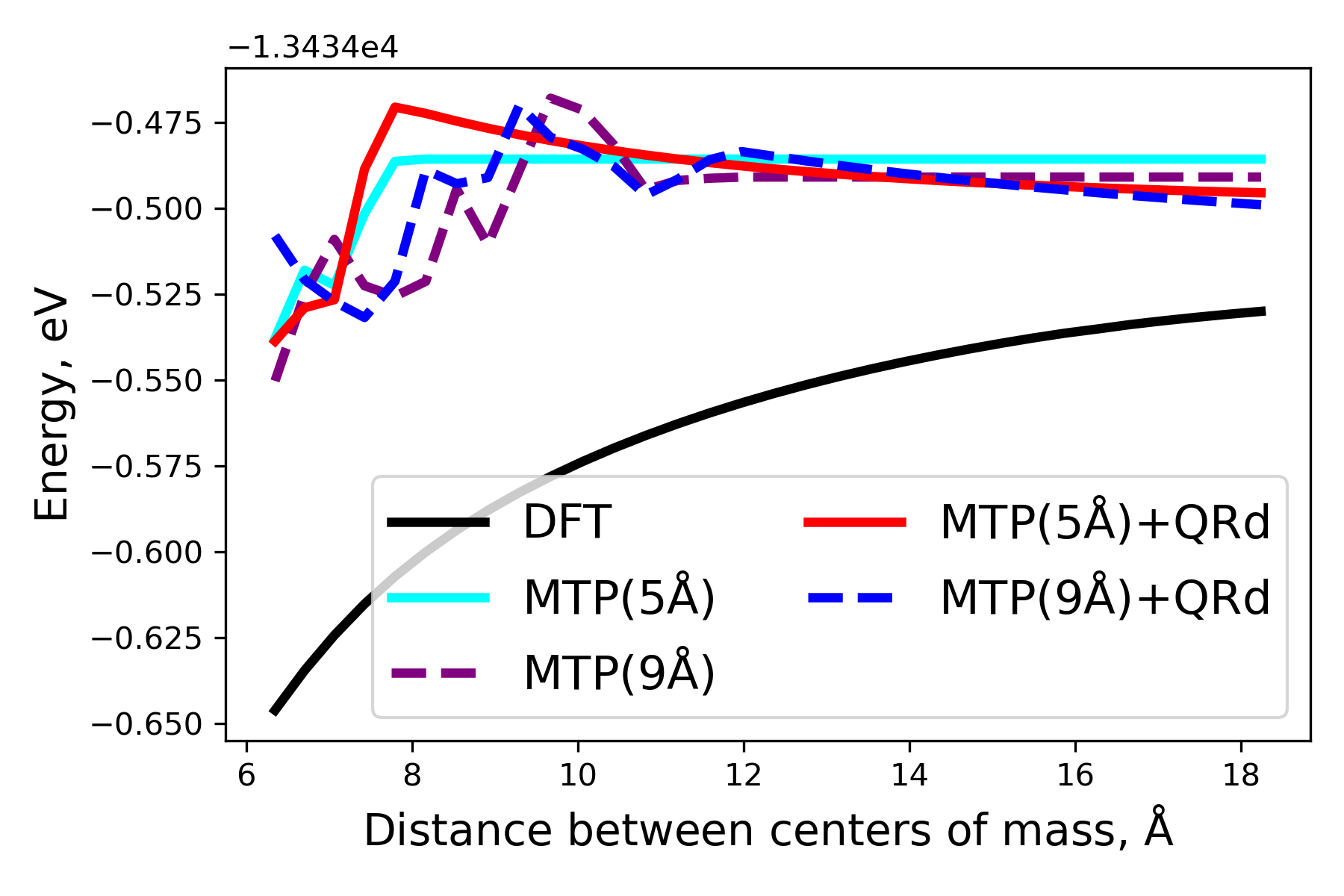}
        \caption{$\rm CH_3COO^-$+4-methylimidazole.}
        \label{fig:binding_curve_imidazole}
    \end{subfigure}
    \caption{Binding curves calculated using DFT and the MLIPs for the $\rm CH_3COO^-$+4-methylphenole~(a) and $\rm CH_3COO^-$+4-methylimidazole~(b) systems.}
    \label{fig:binding_curves_ion_and_mol}
\end{figure}

Interestingly, for the systems containing one charged and one neutral molecule, adding QRd to MTP does not reduce the training energy errors.
From the binding curves shown in Fig.~\ref{fig:binding_curves_ion_and_mol}, we conclude that for the system containing 4-methylimidazole~(Fig.~\ref{fig:binding_curve_imidazole}) the short-range and long-range models are unable to predict the binding curve even qualitatively. We have tried to increase the cutoff radii for both models from 5 to 9 \AA \; but the results did not improve. Nonetheless, the accuracy of the testing binding curve prediction for the system containing 4-methylphenole~(Fig.~\ref{fig:binding_curve_phenole}) is much higher. To compare the accuracy of different MLIPs, we calculated the $L^2$-norm of the differences between the energies predicted using DFT and the MLIPs. These $L^2$-norms are given in Table~\ref{Table:binding_curve_ion_and_mol_errors}.

\begin{table*}[!ht]
\caption{$L^2$-norms of the deviations between binding energies predicted with DFT and the MLIPs.}
\label{Table:binding_curve_ion_and_mol_errors}
\begin{center}
\begin{tabular}{c|c|c}
\hline
\hline
\multirow{2}{*}{System} & \multirow{2}{*}{Model} & \multirow{2}{*}{$\left\Vert E^{\rm Model} - E^{\rm DFT}\right\Vert_{L^2\left(\left[a, b\right]\right)}$,  eV$\cdot \text{\AA}^{0.5}$}\\
& & \\ \hline
\multirow{4}{*}{$\rm CH_3COO^-$+4-methylphenole} & MTP(5\AA) & 0.062 \\ \cline{2-3}
& MTP(5\AA)+QRd & 0.045 \\ \cline {2-3}
& MTP(9\AA) & 0.080 \\ \cline{2-3}
& MTP(9\AA)+QRd & 0.030 \\ \hline 
\hline
\multirow{4}{*}{$\rm CH_3COO^-$+4-methylimidazole} & MTP(5\AA) & 0.27 \\ \cline{2-3}
& MTP(5\AA)+QRd & 0.27 \\ \cline {2-3}
& MTP(9\AA) & 0.25 \\ \cline{2-3}
& MTP(9\AA)+QRd & 0.26\\ \hline 
\hline
\end{tabular}
\end{center}
\end{table*}

$L^2$-norms support previous conclusions: for the system containing 4-methylimidazole adding QRd did not change the accuracy of the binding curve prediction, whereas for the system containing 4-methylphenole adding QRd gave a twofold improvement on average.

Overall, we conclude that QRd is not expressive enough to improve the accuracy of MTP for the $\rm CH_3COO^-$+4-methylimidazole system and more accurate models are required.

\section{Conclusions}

In this study, we explicitly added the long-range electrostatic interactions via Coulomb energy with fixed charges to two machine-learning interatomic potentials: Moment Tensor Potentials and Equivariant Tensor Network Potentials. The long-range models were implemented with the charge redistribution (QRd) technique, which preserves the total charge of a system. We tested the short-range (MTP and ETN) and the long-range (MTP+QRd and ETN+QRd) models on several organic dimers. We compared the results calculated using the MLIPs with the DFT results. First, we demonstrated that the energy root-mean square error (RMSE) calculated with the long-range models (MTP+QRd and ETN+QRd) is 9 times smaller for the CH$_3$COO$^{-}$+CH$_3$COO$^{-}$ system and 4.5 times smaller for the CH$_3$COO$^{-}$+CH$_3$NH$_3^{+}$ system than the energy RMSEs obtained with the short-range models (MTP and ETN), and that the energy RMSE does not depend on the short-range model. Next, we have shown that, both, the MTP+QRd and ETN+QRd models accurately reproduced the DFT binding curves, whereas the MTP and ETN models somehow predicted the DFT binding curves only in a short-range region and failed to capture long-range interactions. Afterwards, we showed that adding the QRd to the MTP for two other systems containing only charged molecules---$\rm C_2H_5NH_3^+$+$\rm C_2H_8N_3^+$ and $\rm CH_3COO^-$+$\rm C_2H_8N_3^+$---significantly reduced the training energy RMSEs and, therefore, MTP+QRd correctly predicted the corresponding binding curves. Next, we demonstrated that using QRd with fixed charges is not enough to reduce the training errors when added to an MTP with a cutoff radius of 5 \AA \;and 9 \AA, respectively, for the dimers consisting of one neutral and one charged molecule~($\rm CH_3COO^-$+4-methylphenole and $\rm CH_3COO^-$+4-methylimidazole). Finally, we showed that MTP+QRd slightly improves on the already acceptable binding curve prediction obtained using the MTP for the system containing 4-methylphenole and, similar to the MTP, fails to predict the binding curve for the system containing 4-methylimidazole.  We therefore conclude that explicitly adding long-range Coulomb interactions with fixed charges to local MLIPs enables for accurately investigating charged organic molecules, especially at long distances between them.

Thus, even a simple long-range model including fixed charges gives a significant improvement of accuracy for the five out of six considered systems and the greatest effect is observed for the systems where all molecules are charged. We expect that significant improvement in the prediction of the binding curves for dimers containing neutral molecules can be achieved by using environment-dependent charges. We are planning to address this in future work.

\section*{Acknowledgments}

This work was in part supported by the Russian Science Foundation (grant number 23-13-00332, https://rscf.ru/project/23-13-00332/).

Max Hodapp gratefully acknowledges the financial support by the Austrian Federal Ministry for Labour and Economy and the National Foundation for Research, Technology and Development and the Christian Doppler Research Association.

\section*{Author declarations}

\subsection*{Conflict of interest}

The authors have no conflicts to disclose.

\subsection*{Author contributions}

\textbf{Dmitry Korogod}: Conceptualization (supporting); Methodology (equal); Software (lead); Formal analysis (lead); Writing - original draft (equal); Writing - review \& editing (equal).
\textbf{Olga Chalykh}: Methodology (supporting); Writing - review \& editing (supporting).
\textbf{Max Hodapp}: Software (supporting); Writing - original draft (equal); Writing - review \& editing (equal).
\textbf{Nikita Rybin}: Methodology (supporting); Writing - review \& editing (supporting).
\textbf{Ivan S. Novikov}: Conceptualization (equal); Methodology (equal); Formal analysis (supporting); Writing - original draft (lead); Writing - review \& editing (equal).
\textbf{Alexander V. Shapeev}: Conceptualization (equal); Methodology (equal); Software (supporting); Writing - review \& editing (supporting).

\section*{Data Availability Statement}

The MLIP-4 code used for the fitting of all MLIPs utilized in this work is publicly available in the repository at https://gitlab.com/ashapeev/mlip-4.

\bibliography{sample}

\begin{thebibliography}{30}%
\makeatletter
\providecommand \@ifxundefined [1]{%
 \@ifx{#1\undefined}
}%
\providecommand \@ifnum [1]{%
 \ifnum #1\expandafter \@firstoftwo
 \else \expandafter \@secondoftwo
 \fi
}%
\providecommand \@ifx [1]{%
 \ifx #1\expandafter \@firstoftwo
 \else \expandafter \@secondoftwo
 \fi
}%
\providecommand \natexlab [1]{#1}%
\providecommand \enquote  [1]{``#1''}%
\providecommand \bibnamefont  [1]{#1}%
\providecommand \bibfnamefont [1]{#1}%
\providecommand \citenamefont [1]{#1}%
\providecommand \href@noop [0]{\@secondoftwo}%
\providecommand \href [0]{\begingroup \@sanitize@url \@href}%
\providecommand \@href[1]{\@@startlink{#1}\@@href}%
\providecommand \@@href[1]{\endgroup#1\@@endlink}%
\providecommand \@sanitize@url [0]{\catcode `\\12\catcode `\$12\catcode `\&12\catcode `\#12\catcode `\^12\catcode `\_12\catcode `\%12\relax}%
\providecommand \@@startlink[1]{}%
\providecommand \@@endlink[0]{}%
\providecommand \url  [0]{\begingroup\@sanitize@url \@url }%
\providecommand \@url [1]{\endgroup\@href {#1}{\urlprefix }}%
\providecommand \urlprefix  [0]{URL }%
\providecommand \Eprint [0]{\href }%
\providecommand \doibase [0]{http://dx.doi.org/}%
\providecommand \selectlanguage [0]{\@gobble}%
\providecommand \bibinfo  [0]{\@secondoftwo}%
\providecommand \bibfield  [0]{\@secondoftwo}%
\providecommand \translation [1]{[#1]}%
\providecommand \BibitemOpen [0]{}%
\providecommand \bibitemStop [0]{}%
\providecommand \bibitemNoStop [0]{.\EOS\space}%
\providecommand \EOS [0]{\spacefactor3000\relax}%
\providecommand \BibitemShut  [1]{\csname bibitem#1\endcsname}%
\let\auto@bib@innerbib\@empty
\bibitem [{\citenamefont {Friederich}\ \emph {et~al.}(2021)\citenamefont {Friederich}, \citenamefont {Häse}, \citenamefont {Proppe},\ and\ \citenamefont {Aspuru-Guzik}}]{MD_MLIPs_REVIEW}%
  \BibitemOpen
  \bibfield  {author} {\bibinfo {author} {\bibfnamefont {P.}~\bibnamefont {Friederich}}, \bibinfo {author} {\bibfnamefont {F.}~\bibnamefont {Häse}}, \bibinfo {author} {\bibfnamefont {J.}~\bibnamefont {Proppe}}, \ and\ \bibinfo {author} {\bibfnamefont {A.}~\bibnamefont {Aspuru-Guzik}},\ }\href@noop {} {\bibfield  {journal} {\bibinfo  {journal} {Nature Materials}\ }\textbf {\bibinfo {volume} {20}},\ \bibinfo {pages} {750} (\bibinfo {year} {2021})}\BibitemShut {NoStop}%
\bibitem [{\citenamefont {Behler}\ and\ \citenamefont {Parrinello}(2007)}]{NNP}%
  \BibitemOpen
  \bibfield  {author} {\bibinfo {author} {\bibfnamefont {J.}~\bibnamefont {Behler}}\ and\ \bibinfo {author} {\bibfnamefont {M.}~\bibnamefont {Parrinello}},\ }\href@noop {} {\bibfield  {journal} {\bibinfo  {journal} {Phys. Rev. Lett.}\ }\textbf {\bibinfo {volume} {98}},\ \bibinfo {pages} {146401} (\bibinfo {year} {2007})}\BibitemShut {NoStop}%
\bibitem [{\citenamefont {Bartok}, \citenamefont {Kondor},\ and\ \citenamefont {Csanyi}(2013)}]{GAP}%
  \BibitemOpen
  \bibfield  {author} {\bibinfo {author} {\bibfnamefont {A.~P.}\ \bibnamefont {Bartok}}, \bibinfo {author} {\bibfnamefont {R.}~\bibnamefont {Kondor}}, \ and\ \bibinfo {author} {\bibfnamefont {G.}~\bibnamefont {Csanyi}},\ }\href@noop {} {\bibfield  {journal} {\bibinfo  {journal} {Phys. Rev. B}\ }\textbf {\bibinfo {volume} {87}},\ \bibinfo {pages} {184115} (\bibinfo {year} {2013})}\BibitemShut {NoStop}%
\bibitem [{\citenamefont {Bart{\'o}k}, \citenamefont {Kondor},\ and\ \citenamefont {Cs{\'a}nyi}(2013)}]{SOAP}%
  \BibitemOpen
  \bibfield  {author} {\bibinfo {author} {\bibfnamefont {A.~P.}\ \bibnamefont {Bart{\'o}k}}, \bibinfo {author} {\bibfnamefont {R.}~\bibnamefont {Kondor}}, \ and\ \bibinfo {author} {\bibfnamefont {G.}~\bibnamefont {Cs{\'a}nyi}},\ }\href@noop {} {\bibfield  {journal} {\bibinfo  {journal} {Physical Review B—Condensed Matter and Materials Physics}\ }\textbf {\bibinfo {volume} {87}},\ \bibinfo {pages} {184115} (\bibinfo {year} {2013})}\BibitemShut {NoStop}%
\bibitem [{\citenamefont {Thompson}\ \emph {et~al.}(2015)\citenamefont {Thompson}, \citenamefont {Swiler}, \citenamefont {Trott}, \citenamefont {Foiles},\ and\ \citenamefont {Tucker}}]{SNAP}%
  \BibitemOpen
  \bibfield  {author} {\bibinfo {author} {\bibfnamefont {A.}~\bibnamefont {Thompson}}, \bibinfo {author} {\bibfnamefont {L.}~\bibnamefont {Swiler}}, \bibinfo {author} {\bibfnamefont {C.}~\bibnamefont {Trott}}, \bibinfo {author} {\bibfnamefont {S.}~\bibnamefont {Foiles}}, \ and\ \bibinfo {author} {\bibfnamefont {G.}~\bibnamefont {Tucker}},\ }\href@noop {} {\bibfield  {journal} {\bibinfo  {journal} {Journal of Computational Physics}\ }\textbf {\bibinfo {volume} {285}},\ \bibinfo {pages} {316} (\bibinfo {year} {2015})}\BibitemShut {NoStop}%
\bibitem [{\citenamefont {Gubaev}, \citenamefont {Podryabinkin},\ and\ \citenamefont {Shapeev}(2018)}]{MTPmulti}%
  \BibitemOpen
  \bibfield  {author} {\bibinfo {author} {\bibfnamefont {K.}~\bibnamefont {Gubaev}}, \bibinfo {author} {\bibfnamefont {E.~V.}\ \bibnamefont {Podryabinkin}}, \ and\ \bibinfo {author} {\bibfnamefont {A.~V.}\ \bibnamefont {Shapeev}},\ }\href@noop {} {\bibfield  {journal} {\bibinfo  {journal} {The Journal of Chemical Physics}\ }\textbf {\bibinfo {volume} {148}},\ \bibinfo {pages} {241727} (\bibinfo {year} {2018})}\BibitemShut {NoStop}%
\bibitem [{\citenamefont {Wang}\ \emph {et~al.}(2018)\citenamefont {Wang}, \citenamefont {Zhang}, \citenamefont {Han},\ and\ \citenamefont {Weinan}}]{DeePMDsr}%
  \BibitemOpen
  \bibfield  {author} {\bibinfo {author} {\bibfnamefont {H.}~\bibnamefont {Wang}}, \bibinfo {author} {\bibfnamefont {L.}~\bibnamefont {Zhang}}, \bibinfo {author} {\bibfnamefont {J.}~\bibnamefont {Han}}, \ and\ \bibinfo {author} {\bibfnamefont {E.}~\bibnamefont {Weinan}},\ }\href@noop {} {\bibfield  {journal} {\bibinfo  {journal} {Computer Physics Communications}\ }\textbf {\bibinfo {volume} {228}},\ \bibinfo {pages} {178} (\bibinfo {year} {2018})}\BibitemShut {NoStop}%
\bibitem [{\citenamefont {Drautz}(2019)}]{ACE}%
  \BibitemOpen
  \bibfield  {author} {\bibinfo {author} {\bibfnamefont {R.}~\bibnamefont {Drautz}},\ }\href {\doibase 10.1103/PhysRevB.99.014104} {\bibfield  {journal} {\bibinfo  {journal} {Phys. Rev. B}\ }\textbf {\bibinfo {volume} {99}},\ \bibinfo {pages} {014104} (\bibinfo {year} {2019})}\BibitemShut {NoStop}%
\bibitem [{\citenamefont {Pun}\ \emph {et~al.}(2019)\citenamefont {Pun}, \citenamefont {Batra}, \citenamefont {Ramprasad},\ and\ \citenamefont {Mishin}}]{PINN}%
  \BibitemOpen
  \bibfield  {author} {\bibinfo {author} {\bibfnamefont {G.~P.}\ \bibnamefont {Pun}}, \bibinfo {author} {\bibfnamefont {R.}~\bibnamefont {Batra}}, \bibinfo {author} {\bibfnamefont {R.}~\bibnamefont {Ramprasad}}, \ and\ \bibinfo {author} {\bibfnamefont {Y.}~\bibnamefont {Mishin}},\ }\href@noop {} {\bibfield  {journal} {\bibinfo  {journal} {Nature communications}\ }\textbf {\bibinfo {volume} {10}},\ \bibinfo {pages} {2339} (\bibinfo {year} {2019})}\BibitemShut {NoStop}%
\bibitem [{\citenamefont {Batzner}\ \emph {et~al.}(2022)\citenamefont {Batzner}, \citenamefont {Musaelian}, \citenamefont {Sun}, \citenamefont {Geiger}, \citenamefont {Mailoa}, \citenamefont {Kornbluth}, \citenamefont {Molinari}, \citenamefont {Smidt},\ and\ \citenamefont {Kozinsky}}]{NeQUIP}%
  \BibitemOpen
  \bibfield  {author} {\bibinfo {author} {\bibfnamefont {S.}~\bibnamefont {Batzner}}, \bibinfo {author} {\bibfnamefont {A.}~\bibnamefont {Musaelian}}, \bibinfo {author} {\bibfnamefont {L.}~\bibnamefont {Sun}}, \bibinfo {author} {\bibfnamefont {M.}~\bibnamefont {Geiger}}, \bibinfo {author} {\bibfnamefont {J.~P.}\ \bibnamefont {Mailoa}}, \bibinfo {author} {\bibfnamefont {M.}~\bibnamefont {Kornbluth}}, \bibinfo {author} {\bibfnamefont {N.}~\bibnamefont {Molinari}}, \bibinfo {author} {\bibfnamefont {T.~E.}\ \bibnamefont {Smidt}}, \ and\ \bibinfo {author} {\bibfnamefont {B.}~\bibnamefont {Kozinsky}},\ }\href@noop {} {\bibfield  {journal} {\bibinfo  {journal} {Nature communications}\ }\textbf {\bibinfo {volume} {13}},\ \bibinfo {pages} {2453} (\bibinfo {year} {2022})}\BibitemShut {NoStop}%
\bibitem [{\citenamefont {Hodapp}\ and\ \citenamefont {Shapeev}(2024)}]{ETN}%
  \BibitemOpen
  \bibfield  {author} {\bibinfo {author} {\bibfnamefont {M.}~\bibnamefont {Hodapp}}\ and\ \bibinfo {author} {\bibfnamefont {A.}~\bibnamefont {Shapeev}},\ }\href@noop {} {\bibfield  {journal} {\bibinfo  {journal} {Machine Learning: Science and Technology}\ }\textbf {\bibinfo {volume} {5}},\ \bibinfo {pages} {035075} (\bibinfo {year} {2024})}\BibitemShut {NoStop}%
\bibitem [{\citenamefont {Artrith}, \citenamefont {Morawietz},\ and\ \citenamefont {Behler}(2011)}]{3GHDNNP}%
  \BibitemOpen
  \bibfield  {author} {\bibinfo {author} {\bibfnamefont {N.}~\bibnamefont {Artrith}}, \bibinfo {author} {\bibfnamefont {T.}~\bibnamefont {Morawietz}}, \ and\ \bibinfo {author} {\bibfnamefont {J.}~\bibnamefont {Behler}},\ }\href@noop {} {\bibfield  {journal} {\bibinfo  {journal} {Phys. Rev. B}\ }\textbf {\bibinfo {volume} {83}},\ \bibinfo {pages} {153101} (\bibinfo {year} {2011})}\BibitemShut {NoStop}%
\bibitem [{\citenamefont {Rappe}\ and\ \citenamefont {Goddard}(1991)}]{QEq}%
  \BibitemOpen
  \bibfield  {author} {\bibinfo {author} {\bibfnamefont {A.~K.}\ \bibnamefont {Rappe}}\ and\ \bibinfo {author} {\bibfnamefont {W.~A.~I.}\ \bibnamefont {Goddard}},\ }\href {\doibase 10.1021/j100161a070} {\bibfield  {journal} {\bibinfo  {journal} {The Journal of Physical Chemistry}\ }\textbf {\bibinfo {volume} {95}},\ \bibinfo {pages} {3358} (\bibinfo {year} {1991})}\BibitemShut {NoStop}%
\bibitem [{\citenamefont {Ghasemi}\ \emph {et~al.}(2015)\citenamefont {Ghasemi}, \citenamefont {Hofstetter}, \citenamefont {Saha},\ and\ \citenamefont {Goedecker}}]{CENT}%
  \BibitemOpen
  \bibfield  {author} {\bibinfo {author} {\bibfnamefont {S.~A.}\ \bibnamefont {Ghasemi}}, \bibinfo {author} {\bibfnamefont {A.}~\bibnamefont {Hofstetter}}, \bibinfo {author} {\bibfnamefont {S.}~\bibnamefont {Saha}}, \ and\ \bibinfo {author} {\bibfnamefont {S.}~\bibnamefont {Goedecker}},\ }\href {\doibase 10.1103/PhysRevB.92.045131} {\bibfield  {journal} {\bibinfo  {journal} {Phys. Rev. B}\ }\textbf {\bibinfo {volume} {92}},\ \bibinfo {pages} {045131} (\bibinfo {year} {2015})}\BibitemShut {NoStop}%
\bibitem [{\citenamefont {Ko}\ \emph {et~al.}(2021)\citenamefont {Ko}, \citenamefont {Finkler}, \citenamefont {Goedecker},\ and\ \citenamefont {Behler}}]{4G-HDNNP}%
  \BibitemOpen
  \bibfield  {author} {\bibinfo {author} {\bibfnamefont {T.~W.}\ \bibnamefont {Ko}}, \bibinfo {author} {\bibfnamefont {J.~A.}\ \bibnamefont {Finkler}}, \bibinfo {author} {\bibfnamefont {S.}~\bibnamefont {Goedecker}}, \ and\ \bibinfo {author} {\bibfnamefont {J.}~\bibnamefont {Behler}},\ }\href {\doibase 10.1021/acs.accounts.0c00689} {\bibfield  {journal} {\bibinfo  {journal} {Accounts of Chemical Research}\ }\textbf {\bibinfo {volume} {54}},\ \bibinfo {pages} {808} (\bibinfo {year} {2021})}\BibitemShut {NoStop}%
\bibitem [{\citenamefont {Ko}\ \emph {et~al.}(2023)\citenamefont {Ko}, \citenamefont {Finkler}, \citenamefont {Goedecker},\ and\ \citenamefont {Behler}}]{ee4GHDNNP}%
  \BibitemOpen
  \bibfield  {author} {\bibinfo {author} {\bibfnamefont {T.~W.}\ \bibnamefont {Ko}}, \bibinfo {author} {\bibfnamefont {J.~A.}\ \bibnamefont {Finkler}}, \bibinfo {author} {\bibfnamefont {S.}~\bibnamefont {Goedecker}}, \ and\ \bibinfo {author} {\bibfnamefont {J.}~\bibnamefont {Behler}},\ }\href@noop {} {\bibfield  {journal} {\bibinfo  {journal} {Journal of Chemical Theory and Computation}\ }\textbf {\bibinfo {volume} {19}},\ \bibinfo {pages} {3567} (\bibinfo {year} {2023})}\BibitemShut {NoStop}%
\bibitem [{\citenamefont {Yue}\ \emph {et~al.}(2021)\citenamefont {Yue}, \citenamefont {Muniz}, \citenamefont {Calegari~Andrade}, \citenamefont {Zhang}, \citenamefont {Car},\ and\ \citenamefont {Panagiotopoulos}}]{DeePMDlr}%
  \BibitemOpen
  \bibfield  {author} {\bibinfo {author} {\bibfnamefont {S.}~\bibnamefont {Yue}}, \bibinfo {author} {\bibfnamefont {M.~C.}\ \bibnamefont {Muniz}}, \bibinfo {author} {\bibfnamefont {M.~F.}\ \bibnamefont {Calegari~Andrade}}, \bibinfo {author} {\bibfnamefont {L.}~\bibnamefont {Zhang}}, \bibinfo {author} {\bibfnamefont {R.}~\bibnamefont {Car}}, \ and\ \bibinfo {author} {\bibfnamefont {A.~Z.}\ \bibnamefont {Panagiotopoulos}},\ }\href@noop {} {\bibfield  {journal} {\bibinfo  {journal} {The Journal of Chemical Physics}\ }\textbf {\bibinfo {volume} {154}} (\bibinfo {year} {2021})}\BibitemShut {NoStop}%
\bibitem [{\citenamefont {Zhang}\ \emph {et~al.}(2022)\citenamefont {Zhang}, \citenamefont {Wang}, \citenamefont {Muniz}, \citenamefont {Panagiotopoulos}, \citenamefont {Car},\ and\ \citenamefont {E}}]{DPLR}%
  \BibitemOpen
  \bibfield  {author} {\bibinfo {author} {\bibfnamefont {L.}~\bibnamefont {Zhang}}, \bibinfo {author} {\bibfnamefont {H.}~\bibnamefont {Wang}}, \bibinfo {author} {\bibfnamefont {M.~C.}\ \bibnamefont {Muniz}}, \bibinfo {author} {\bibfnamefont {A.~Z.}\ \bibnamefont {Panagiotopoulos}}, \bibinfo {author} {\bibfnamefont {R.}~\bibnamefont {Car}}, \ and\ \bibinfo {author} {\bibfnamefont {W.}~\bibnamefont {E}},\ }\href {\doibase 10.1063/5.0083669} {\bibfield  {journal} {\bibinfo  {journal} {The Journal of Chemical Physics}\ }\textbf {\bibinfo {volume} {156}},\ \bibinfo {pages} {124107} (\bibinfo {year} {2022})}\BibitemShut {NoStop}%
\bibitem [{\citenamefont {Marzari}\ \emph {et~al.}(2012)\citenamefont {Marzari}, \citenamefont {Mostofi}, \citenamefont {Yates}, \citenamefont {Souza},\ and\ \citenamefont {Vanderbilt}}]{WannierCenter}%
  \BibitemOpen
  \bibfield  {author} {\bibinfo {author} {\bibfnamefont {N.}~\bibnamefont {Marzari}}, \bibinfo {author} {\bibfnamefont {A.~A.}\ \bibnamefont {Mostofi}}, \bibinfo {author} {\bibfnamefont {J.~R.}\ \bibnamefont {Yates}}, \bibinfo {author} {\bibfnamefont {I.}~\bibnamefont {Souza}}, \ and\ \bibinfo {author} {\bibfnamefont {D.}~\bibnamefont {Vanderbilt}},\ }\href {\doibase 10.1103/RevModPhys.84.1419} {\bibfield  {journal} {\bibinfo  {journal} {Rev. Mod. Phys.}\ }\textbf {\bibinfo {volume} {84}},\ \bibinfo {pages} {1419} (\bibinfo {year} {2012})}\BibitemShut {NoStop}%
\bibitem [{\citenamefont {Grisafi}\ and\ \citenamefont {Ceriotti}(2019)}]{LODE}%
  \BibitemOpen
  \bibfield  {author} {\bibinfo {author} {\bibfnamefont {A.}~\bibnamefont {Grisafi}}\ and\ \bibinfo {author} {\bibfnamefont {M.}~\bibnamefont {Ceriotti}},\ }\href {\doibase 10.1063/1.5128375} {\bibfield  {journal} {\bibinfo  {journal} {The Journal of Chemical Physics}\ }\textbf {\bibinfo {volume} {151}},\ \bibinfo {pages} {204105} (\bibinfo {year} {2019})}\BibitemShut {NoStop}%
\bibitem [{\citenamefont {Cheng}(2024)}]{CACE}%
  \BibitemOpen
  \bibfield  {author} {\bibinfo {author} {\bibfnamefont {B.}~\bibnamefont {Cheng}},\ }\href@noop {} {\bibfield  {journal} {\bibinfo  {journal} {npj Computational Materials}\ }\textbf {\bibinfo {volume} {10}},\ \bibinfo {pages} {157} (\bibinfo {year} {2024})}\BibitemShut {NoStop}%
\bibitem [{\citenamefont {Cheng}(2025)}]{LatentEwald}%
  \BibitemOpen
  \bibfield  {author} {\bibinfo {author} {\bibfnamefont {B.}~\bibnamefont {Cheng}},\ }\href@noop {} {\bibfield  {journal} {\bibinfo  {journal} {npj Computational Materials}\ }\textbf {\bibinfo {volume} {11}},\ \bibinfo {pages} {80} (\bibinfo {year} {2025})}\BibitemShut {NoStop}%
\bibitem [{\citenamefont {Novikov}\ and\ \citenamefont {Shapeev}(2019)}]{MTP+QEq}%
  \BibitemOpen
  \bibfield  {author} {\bibinfo {author} {\bibfnamefont {I.~S.}\ \bibnamefont {Novikov}}\ and\ \bibinfo {author} {\bibfnamefont {A.~V.}\ \bibnamefont {Shapeev}},\ }\href {\doibase 10.1016/j.mtcomm.2018.11.008} {\bibfield  {journal} {\bibinfo  {journal} {Materials Today Communications}\ }\textbf {\bibinfo {volume} {18}},\ \bibinfo {pages} {74} (\bibinfo {year} {2019})}\BibitemShut {NoStop}%
\bibitem [{\citenamefont {Oseledets}(2011)}]{TT}%
  \BibitemOpen
  \bibfield  {author} {\bibinfo {author} {\bibfnamefont {I.~V.}\ \bibnamefont {Oseledets}},\ }\href@noop {} {\bibfield  {journal} {\bibinfo  {journal} {SIAM Journal on Scientific Computing}\ }\textbf {\bibinfo {volume} {33}},\ \bibinfo {pages} {2295} (\bibinfo {year} {2011})}\BibitemShut {NoStop}%
\bibitem [{\citenamefont {Stukowski}(2010)}]{OVITO}%
  \BibitemOpen
  \bibfield  {author} {\bibinfo {author} {\bibfnamefont {A.}~\bibnamefont {Stukowski}},\ }\href@noop {} {\bibfield  {journal} {\bibinfo  {journal} {Modelling and Simulation in Materials Science and Engineering}\ }\textbf {\bibinfo {volume} {{18}}} (\bibinfo {year} {{2010}})}\BibitemShut {NoStop}%
\bibitem [{\citenamefont {Kuhne}\ \emph {et~al.}(2020)\citenamefont {Kuhne}, \citenamefont {Iannuzzi}, \citenamefont {Del~Ben}, \citenamefont {Rybkin}, \citenamefont {Seewald}, \citenamefont {Stein}, \citenamefont {Laino}, \citenamefont {Khaliullin}, \citenamefont {Schutt}, \citenamefont {Schiffmann}, \citenamefont {Golze}, \citenamefont {Wilhelm}, \citenamefont {Chulkov}, \citenamefont {Bani-Hashemian}, \citenamefont {Weber}, \citenamefont {Borstnik}, \citenamefont {Taillefumier}, \citenamefont {Jakobovits}, \citenamefont {Lazzaro}, \citenamefont {Pabst}, \citenamefont {Müller}, \citenamefont {Schade}, \citenamefont {Guidon}, \citenamefont {Andermatt}, \citenamefont {Holmberg}, \citenamefont {Schenter}, \citenamefont {Hehn}, \citenamefont {Bussy}, \citenamefont {Belleflamme}, \citenamefont {Tabacchi}, \citenamefont {Glob}, \citenamefont {Lass}, \citenamefont {Bethune}, \citenamefont {Mundy}, \citenamefont {Plessl}, \citenamefont {Watkins}, \citenamefont {VandeVondele}, \citenamefont {Krack},\ and\
  \citenamefont {Hutter}}]{CP2K}%
  \BibitemOpen
  \bibfield  {author} {\bibinfo {author} {\bibfnamefont {T.~D.}\ \bibnamefont {Kuhne}}, \bibinfo {author} {\bibfnamefont {M.}~\bibnamefont {Iannuzzi}}, \bibinfo {author} {\bibfnamefont {M.}~\bibnamefont {Del~Ben}}, \bibinfo {author} {\bibfnamefont {V.~V.}\ \bibnamefont {Rybkin}}, \bibinfo {author} {\bibfnamefont {P.}~\bibnamefont {Seewald}}, \bibinfo {author} {\bibfnamefont {F.}~\bibnamefont {Stein}}, \bibinfo {author} {\bibfnamefont {T.}~\bibnamefont {Laino}}, \bibinfo {author} {\bibfnamefont {R.~Z.}\ \bibnamefont {Khaliullin}}, \bibinfo {author} {\bibfnamefont {O.}~\bibnamefont {Schutt}}, \bibinfo {author} {\bibfnamefont {F.}~\bibnamefont {Schiffmann}}, \bibinfo {author} {\bibfnamefont {D.}~\bibnamefont {Golze}}, \bibinfo {author} {\bibfnamefont {J.}~\bibnamefont {Wilhelm}}, \bibinfo {author} {\bibfnamefont {S.}~\bibnamefont {Chulkov}}, \bibinfo {author} {\bibfnamefont {M.~H.}\ \bibnamefont {Bani-Hashemian}}, \bibinfo {author} {\bibfnamefont {V.}~\bibnamefont {Weber}}, \bibinfo {author} {\bibfnamefont
  {U.}~\bibnamefont {Borstnik}}, \bibinfo {author} {\bibfnamefont {M.}~\bibnamefont {Taillefumier}}, \bibinfo {author} {\bibfnamefont {A.~S.}\ \bibnamefont {Jakobovits}}, \bibinfo {author} {\bibfnamefont {A.}~\bibnamefont {Lazzaro}}, \bibinfo {author} {\bibfnamefont {H.}~\bibnamefont {Pabst}}, \bibinfo {author} {\bibfnamefont {T.}~\bibnamefont {Müller}}, \bibinfo {author} {\bibfnamefont {R.}~\bibnamefont {Schade}}, \bibinfo {author} {\bibfnamefont {M.}~\bibnamefont {Guidon}}, \bibinfo {author} {\bibfnamefont {S.}~\bibnamefont {Andermatt}}, \bibinfo {author} {\bibfnamefont {N.}~\bibnamefont {Holmberg}}, \bibinfo {author} {\bibfnamefont {G.~K.}\ \bibnamefont {Schenter}}, \bibinfo {author} {\bibfnamefont {A.}~\bibnamefont {Hehn}}, \bibinfo {author} {\bibfnamefont {A.}~\bibnamefont {Bussy}}, \bibinfo {author} {\bibfnamefont {F.}~\bibnamefont {Belleflamme}}, \bibinfo {author} {\bibfnamefont {G.}~\bibnamefont {Tabacchi}}, \bibinfo {author} {\bibfnamefont {A.}~\bibnamefont {Glob}}, \bibinfo {author} {\bibfnamefont
  {M.}~\bibnamefont {Lass}}, \bibinfo {author} {\bibfnamefont {I.}~\bibnamefont {Bethune}}, \bibinfo {author} {\bibfnamefont {C.~J.}\ \bibnamefont {Mundy}}, \bibinfo {author} {\bibfnamefont {C.}~\bibnamefont {Plessl}}, \bibinfo {author} {\bibfnamefont {M.}~\bibnamefont {Watkins}}, \bibinfo {author} {\bibfnamefont {J.}~\bibnamefont {VandeVondele}}, \bibinfo {author} {\bibfnamefont {M.}~\bibnamefont {Krack}}, \ and\ \bibinfo {author} {\bibfnamefont {J.}~\bibnamefont {Hutter}},\ }\href@noop {} {\bibfield  {journal} {\bibinfo  {journal} {The Journal of Chemical Physics}\ }\textbf {\bibinfo {volume} {152}},\ \bibinfo {pages} {194103} (\bibinfo {year} {2020})}\BibitemShut {NoStop}%
\bibitem [{\citenamefont {Grimme}\ \emph {et~al.}(2010)\citenamefont {Grimme}, \citenamefont {Antony}, \citenamefont {Ehrlich},\ and\ \citenamefont {Krieg}}]{GrimmeD3}%
  \BibitemOpen
  \bibfield  {author} {\bibinfo {author} {\bibfnamefont {S.}~\bibnamefont {Grimme}}, \bibinfo {author} {\bibfnamefont {J.}~\bibnamefont {Antony}}, \bibinfo {author} {\bibfnamefont {S.}~\bibnamefont {Ehrlich}}, \ and\ \bibinfo {author} {\bibfnamefont {H.}~\bibnamefont {Krieg}},\ }\href@noop {} {\bibfield  {journal} {\bibinfo  {journal} {The Journal of Chemical Physics}\ }\textbf {\bibinfo {volume} {132}},\ \bibinfo {pages} {154104} (\bibinfo {year} {2010})}\BibitemShut {NoStop}%
\bibitem [{\citenamefont {Mulliken}(1955)}]{MullikenCharges}%
  \BibitemOpen
  \bibfield  {author} {\bibinfo {author} {\bibfnamefont {R.~S.}\ \bibnamefont {Mulliken}},\ }\href {\doibase 10.1063/1.1740588} {\bibfield  {journal} {\bibinfo  {journal} {The Journal of Chemical Physics}\ }\textbf {\bibinfo {volume} {23}},\ \bibinfo {pages} {1833} (\bibinfo {year} {1955})}\BibitemShut {NoStop}%
\bibitem [{FHI(2009)}]{FHI-AIMS}%
  \BibitemOpen
  \href@noop {} {\bibfield  {journal} {\bibinfo  {journal} {Computer Physics Communications}\ }\textbf {\bibinfo {volume} {180}},\ \bibinfo {pages} {2175} (\bibinfo {year} {2009})}\BibitemShut {NoStop}%
\bibitem [{\citenamefont {Tkatchenko}\ and\ \citenamefont {Scheffler}(2009)}]{dispersion_TS}%
  \BibitemOpen
  \bibfield  {author} {\bibinfo {author} {\bibfnamefont {A.}~\bibnamefont {Tkatchenko}}\ and\ \bibinfo {author} {\bibfnamefont {M.}~\bibnamefont {Scheffler}},\ }\href {\doibase 10.1103/PhysRevLett.102.073005} {\bibfield  {journal} {\bibinfo  {journal} {Phys. Rev. Lett.}\ }\textbf {\bibinfo {volume} {102}},\ \bibinfo {pages} {073005} (\bibinfo {year} {2009})}\BibitemShut {NoStop}%
\end{thebibliography}%

\end{document}